\documentclass[journal]{IEEEtran}

\ifCLASSINFOpdf
\else
\fi

\usepackage{bm}
\usepackage{cite}
\usepackage{amsmath,amssymb,amsfonts}
\usepackage{booktabs}
\usepackage{graphicx,xcolor}
\usepackage{textcomp}
\usepackage{stfloats}
\usepackage{booktabs}
\usepackage{setspace}
\usepackage{array}
\usepackage[linesnumbered,ruled]{algorithm2e}
\usepackage[caption=false,font=normalsize,labelfont=sf,textfont=sf]{subfig}
\def\BibTeX{{\rm B\kern-.05em{\sc i\kern-.025em b}\kern-.08em  T\kern-.1667em\lower.7ex\hbox{E}\kern-.125emX}}
\usepackage{balance}
\usepackage{algorithmic}
\usepackage{makecell}

\begin{document}

\title{Combinatorial-restless-bandit-based Transmitter-Receiver Online Selection for Distributed MIMO Radars With Non-Stationary Channels}
\author{Yuhang Hao, Zengfu Wang, Jing Fu, Xianglong Bai, Can Li and Quan Pan
\thanks{This work was in part supported by the National Natural Science Foundation of China~(grant no. U21B2008, 62233014).}
\thanks{Yuhang Hao, Zengfu Wang, Xianglong Bai, Can Li, Quan Pan are with the School of Automation, Northwestern Polytechnical University, and the Key Laboratory of Information Fusion Technology, Ministry of Education, Xi'an, Shaanxi, 710072, China.
Zengfu Wang is also with the Research \& Development Institute of Northwestern Polytechnical University in Shenzhen, Shenzhen 518057, China.
Jing Fu is with School of Engineering, RMIT University, Melbourne, VIC, 3000, Australia.
E-mail: (yuhanghao@mail.nwpu.edu.cn; wangzengfu@nwpu.edu.cn;
jing.fu@rmit.edu.au; baixianglong@mail.nwpu.edu.cn; licanvol@163.com; quanpan@nwpu.edu.cn). 
(Corresponding author: Zengfu Wang.)
}
}



\IEEEtitleabstractindextext{%
\begin{abstract}
We track moving targets with a distributed multiple-input multiple-output (MIMO) radar, for which the transmitters and receivers are appropriately paired and selected with a limited number of radar stations.
We aim to maximize the sum of the signal-to-interference-plus-noise ratios (SINRs) of all the targets by sensibly selecting the transmitter-receiver pairs during the tracking period.
A key is to model the optimization problem of selecting the transmitter-receiver pairs by a restless multi-armed bandit (RMAB) model that is able to formulate the time-varying signals of the transceiver channels whenever the channels are being probed or not.
We regard the estimated mean reward~(i.e., SINR) as the state of an arm.
If an arm is probed, the estimated mean reward of the arm is the weighted sum of the observed reward and the predicted mean reward; otherwise, it is the predicted mean reward.
We associate the predicted mean reward with the estimated mean reward at the previous time slot and the state of the target, which is estimated via the interacting multiple model-unscented Kalman filter (IMM-UKF). 
The optimized selection of transmitter-receiver pairs at each time is accomplished by using Binary Particle Swarm Optimization (BPSO) based on indexes of arms, each of which is designed by the upper confidence bound~(UCB1) algorithm.
Above all, a multi-group combinatorial-restless-bandit technique 
taking into account of different combinations of transmitters and receivers and the closed-loop scheme between transmitter-receiver pair selection and target state estimation, 
namely MG-CRB-CL, is developed to achieve a near-optimal selection strategy and improve multi-target tracking performance.
Simulation results for different scenarios are provided to verify the effectiveness and superior performance of our MG-CRB-CL algorithm.
\end{abstract}

}

\maketitle

\IEEEdisplaynontitleabstractindextext

%
\IEEEpeerreviewmaketitle

\section{Introduction}
\IEEEPARstart{I}n a distributed multiple-input multiple-output~(MIMO) radar system, each transmitter-receiver~(TX-RX) pair forms an independent channel for illuminating and observing targets.
The TX-RX pairs can be combined to increase the additional spatial diversity~\cite{haimovich2007mimo,niu2012target}. 
The schemes of TX-RX pair optimal selection or the adaptive channel assignment algorithms, which can simultaneously obtain low computational cost and maintain preset performance, have drawn extensive attention in recent years~\cite{wang2013parametric,nosrati2017receiver,ajorloo2020antenna,wang2014reconfigurable,yan2022radar}. 

The existing work on MIMO radar resource management can be mainly divided into two types of optimization models.
The first one minimizes the scheduling cost or the number of selected TX-RX pairs such that the predetermined performance requirement is guaranteed \cite{lu2019adaptive,zhang2020joint_sensors,shi2020low,deligiannis2016game}.
In the second type of optimization problem model, given a resource budget for a MIMO radar system, a certain amount of resources is selected to maximize the multiple target tracking performance.
In this paper, we focus on the TX-RX pairs selection for the first type in order to achieve the best tracking performance for multiple moving targets.

In general, the TX-RX pairs selection problem for a distributed MIMO radar system is NP-hard~\cite{nosrati2017receiver,godrich2011sensor}, and appropriate algorithms that do not consume an excessively large amount of computational power have been explored.
In \cite{nosrati2017receiver}, a factored selection scheme, where TXs and RXs were selected separately, was established to minimize the spatial correlation
coefficient while reducing the number of solution spaces.
In \cite{godrich2011sensor},
the antenna subset selection in a distributed multi-radar system was modeled as a knapsack problem with Cram\'{e}r-Rao lower bound (CRLB) being the performance metric and a multi-start local search algorithm was proposed, 
where a greedy strategy was employed in each search of the multi-start knapsack tree. 
In \cite{tharmarasa2007large,tharmarasa2007pcrlb},
the convex optimization method followed by greedy local search
and the posterior CRLB (PCRLB) \cite{yan2015simultaneous} were used to find the optimal subsets of the large-scale sensor in two circumstances, 
including the known and fixed number of targets and the unknown and time-varying number of targets in the surveillance region.
In \cite{zhang2022dynamic}, the PCRLB of joint direction-of-arrival (DOA) and Doppler estimate was used as the optimization criterion, and a discrete particle swarm optimization algorithm (PSO) was proposed to approximate optimality. 
Similarly, a modified PSO was developed in \cite{zhang2020finite}, where the hierarchy penalty functions, the crossover operation, and the mutation operation were introduced to enhance the global exploration ability of the PSO. 
In \cite{zhang2020antenna}, an efficient algorithm, which integrated the convex relaxation technique with the local search, was developed to solve the antenna selection problem. 
In \cite{li2019radar}, the semidefinite relaxation (SDR) method was used to solve the large-scale radar selection problem.

In some sense, antenna selection and channel selection problems can be transformed into each other, 
if the latter problem satisfies the constraint that the selected channels come from the same selected TXs and RXs.
In \cite{dai2022adaptive}, two adaptive channel assignment schemes were implemented through non-linear integer programming for maneuvering target tracking. The predicted conditional CRLB, which serves as the metric, was calculated according to the predicted probability density function of the target state and the predictive model probability from the interacting multiple model~(IMM) algorithm. 
For multi-target tracking, a transmit channel assignment problem was considered in \cite{bogdanovic2018target} with a game-theoretic perspective.
In \cite{li2022transmit, yi2020resource,zhang2020joint,xie2017joint}, joint antenna selection was considered, in conjunction with appropriate beam resources or power allocations, for multi-target tracking. These studies typically treated resources separately and used iterative algorithms or multi-step programming approaches to solve resource scheduling problems based on the corresponding CRLB family objective function. 
After obtaining the target dynamic state at the current time, 
a closed-loop feedback system is established to predict the PCRLB and to guide the resource allocation for the next time slot~ \cite{zhang2022dynamic,zhang2020antenna,dai2022adaptive,yi2020resource,zhang2020joint,xie2017joint}.

In the above-mentioned work, 
the channel states of MIMO radar systems were assumed stationary with known distributions over time, and the performance metrics can be computed through deterministic equations.
However, in many real-world applications with rapidly changing environment, the channel states of signal propagation, e.g. SINR, are non-stationary with unknown distributions \cite{pulkkinen2020reinforcement,mukherjee2012learning, kuai2019transmit}. The stationary assumption may incur significant performance degradation, even in the closed-loop framework.

For a channel selection problem without assuming known distributions, the exploration-exploitation trade-off persists for a long-term optimization objective~\cite{pulkkinen2020reinforcement}. 
The moving targets, the time-varying states, and environment conditions for each channel prevent conventional algorithms, such as genetic algorithm (GA) or PSO algorithm, from being directly applied here in an online manner. To tackle this online learning problem, the multi-armed bandit (MAB) model has been exploited, where an agent decides on a bandit process from many parallel ones and tries to find an optimal future action based on the information collected from past observations~\cite{mukherjee2012learning}.
The MAB model has been widely studied in modeling unknown circumstances and reinforcement learning~\cite{gittins2011multi,lattimore2020bandit}. 
In standard MAB, each of the bandit processes is assumed to be independent and identically distributed according to an unknown distribution that characterizes each machine.
If selected in a decision epoch, the bandit process gains an instantaneous reward and the mean reward of the unknown distribution is updated based on it; otherwise, no reward or observation is obtained.
The standard MAB problem aims to maximize the cumulative reward (or minimize the cost) subject to the constraint that only one bandit process is selected at each decision epoch. 
The combinatorial bandit technique was developed and referred to as the combinatorial MAB (CMAB) technique in \cite{chen2016combinatorial} to tackle simultaneous selections of multiple bandit processes in a special case of the restless multi-armed bandit (RMAB) model proposed in~\cite{whittle1988restless}.
The CMAB technique treats each basic project as an arm and selects multiple arms together to make up a combination, also called a super arm, to complete scheduling tasks~\cite{chen2016combinatorial}.
In \cite{pulkkinen2020reinforcement}, the TX-RX selection problems with stationary or moving targets were formulated and solved by the CMAB technique. Nonetheless, the expected SINR reward was assumed to be frozen when the corresponding channel was not observed. 
It cannot model our moving targets for which the underlying target states and the associated SINRs evolve all the time. Also, simulation results showed that the upper confidence bound (UCB1) did not effectively adapt to the non-stationary channel.
In \cite{mukherjee2012learning, kuai2019transmit}, the CMAB technique was used to solve the antenna selection problem via online antenna performance learning, where the reward distribution of each arm corresponding to an antenna subset was identical all over the time. This assumption is not appropriate for our work with moving targets, where the mean reward of each channel is time-varying.

There have been several attempts~\cite{besbes2014stochastic,besbes2015non,whittle1988restless,la2006optimal,nino2022multi} to consider time-varying reward distributions, which are intrinsic in the MAB model~\cite{gittins2011multi}. 
The work in \cite{besbes2014stochastic} formulated a general class of non-stationary reward structures and established a main constraint that the evolution of the mean rewards is bounded by the variation budget. 
In particular, in \cite{whittle1988restless,la2006optimal,nino2022multi},
the restless bandits model was introduced, which defined the states (associated with reward distributions) of arms change in each step according to an arbitrary, yet known, stochastic process. The state representation of each arm, as established in this work, may be valuable for our research as it enables us to capture the transition of the mean reward of arms over time.

In this paper, we consider non-stationary channels and aim to improve the target tracking performance, which can be evaluated by the SINR value of the selected radar system and the root mean square error (RMSE) of target state estimation. 
Here, the SINR model is related to the target slant range and the scatter angles. 
The optimization problem of selecting the TX-RX pairs is substantially complicated by the unknown and time-varying SINR mean reward (expected reward) of each channel. 
Each TX-RX pair and its associated channel is modeled through the restless bandit technique with the non-stationary propagation environment being learned by the UCB1 algorithm.
In order to approximate the expected SINR rewards more accurately, we define the estimated SINR mean reward as the state of each channel in the restless bandits model, where the transition functions are defined based on the estimated target states, the predicted target states, and the radar equation. 
The IMM-UKF technique is used to update the target states with measurements obtained by the given set of selected TX-RX pairs and to solve the maneuvering target tracking problem. 
To this end, we formulate the selection of TX-RX pairs as a combinatorial-restless-bandit~(CRB) problem, where combinatorial arms are selected by Binary PSO (BPSO) based on the index value of each arm by UCB1.
Based on the CRB and IMM-UKF techniques, we propose a closed-loop optimization framework in the manner of an RMAB process and consider a long-term objective with an algorithm proposed to sensibly selects the TX-RX pairs at each decision epoch.
Considering multiple target tracking with different reward distributions of channels of the MIMO radar system, we establish a multi-group CRB (MG-CRB) model, where each group represents all corresponding bandit processes associated with a target.
We refer to the proposed algorithm as the MG-CRB closed-loop (MG-CRB-CL) algorithm.
Through extensive simulation results, our proposed algorithm is numerically demonstrated to outperform UCB1 and other baseline algorithms with respect to SINR and RMSE.
In summary, the main contributions of this paper are three-fold.
\begin{enumerate}
\item 
We model the TX-RX selection problem in a distributed MIMO radar system in the manner of an RMAB process, 
where the SINR states for the unobserved (unselected) TX-RX pairs potentially transit to the predicted states.
We adapt the CMAB technique to the more practical case with time-varying reward distributions and propose the MG-CRB-CL algorithm for the TX-RX selection problem considering the trade-off between exploitation and exploration.
We improve the approximation method proposed in \cite{pulkkinen2020reinforcement} for the expected SINR reward by predicting the target dynamic states. 
Given the set of the selected TX-RX pairs and their observations of SINR, the channel SINR states are approximated by the weighted-fusing measurements and the prediction of the dynamic state. The SINR states of the unobserved (unselected) TX-RX pairs take the predicted value.
We establish the MG-CRB model to formulate the updated SINR states of the bandit processes associated with multiple targets. 
\item 
The TX-RX selection problem in a distributed MIMO radar system is significantly complicated by the dynamics of the multiple targets and the large number of combinatorial subsets of TX-RXs, where optimal solutions are intractable in general. We apply the BPSO algorithm to our TX-RX selection problem, where the transmitters and the receivers are arrayed in one particle vector, of which each binary bit is used to indicate whether the radar is selected or not, for approximating the global optimum.

\item 
For the maneuvering targets, we adapt IMM-UKF to estimate the target dynamic state when the set of the selected TX-RX pairs is given. 
Based on the MG-CRB and IMM-UKF techniques, 
we establish a closed-loop optimization framework, where the 
accurate estimation of the SINR states is enabled by tracking the target dynamic states.
Meanwhile, as the radar measurement errors are affected by the true SINR, the appropriate selection of TX-RX pairs significantly improves the accuracy, evaluated by RMSE, of tracking targets. This improvement on measurement accuracy leads to higher quality of tracking targets and results in a more accurate estimation of the SINR states.
\end{enumerate}

The remainder of this paper is organized as follows. 
In Section \ref{sec:problemformualtion}, the MIMO radar signal model, SINR model, the target dynamic model, and the measurement model are defined, and the TX-RX selection problem is formulated.
The key is to model the multi-target tracking in an RMAB manner and propose the MG-CRB technique to approximate the global optimum. 
In Section \ref{sec:solutions}, the channel selection scheme is developed in the MG-CRB model. The BPSO approach is introduced to achieve the combinatorial optimum of sub-problems associated with the TX-RX pairs based on the SINR state approximations of each TX-RX pair.
Also, the IMM-UKF method with distributed MIMO radar system measurements is derived from distributed fusion method for target tracking and an efficient closed-loop framework is established. 
In Section \ref{sec:simulation}, three simulation scenarios are considered, and the simulation results demonstrate the effectiveness and amenability of the MG-CRB-CL algorithm in the tested scenarios. 
Section \ref{sec:conclusion} concludes this paper.
 

\section{Problem Formulation}\label{sec:problemformualtion}
In this section, we formulate the MIMO radar, SINR, target dynamics, and radar measurement models in a distributed MIMO radar system, which are essential in describing the signal and filtering process of the radar system.
We formulate each channel associated with a TX-RX pair as a restless bandit process 
\cite{gittins2011multi}, of which the state represents the mean SINR reward of the channel. The restless bandit processes for all the TX-RX pairs comprise an RMAB problem~\cite{whittle1988restless}.
We alternatively refer to such a restless bandit process as an arm of an RMAB problem. 
In particular, in the definition of the RMAB problem, each bandit process keeps evolving, potentially into different states (that is, the SINR state of the associated channel is time-varying), even if the corresponding TX-RX pair is not selected and the channel is not observed/detected. The RMAB model is appropriate for real-world applications with moving targets.
We further propose the multi-group (MG) structure in conjunction with the RMAB to capture the dynamic features of the multiple moving targets.

\subsection{MIMO radar modeling}\label{subsec:II-A}
%
Consider a distributed MIMO radar system consisting of $M$ transmitters and $N$ receivers, which is used to track $K$ targets.
The transmitter $m=1,2,\ldots,M$ and the receiver $n=1,2,\ldots,N$ are located in $(x_{m_{Tx}}, y_{m_{Tx}})$ and $(x_{n_{Rx}}, y_{n_{Rx}})$, respectively, in an \textit{x-y} plane.
Let $\bm{\Gamma}_T$ and $\bm{\Gamma}_R$ represent the sets of the selected transmitters and receivers, respectively, for which $| \bm{\Gamma}_T |=M_s$, $| \bm{\Gamma}_R |=N_s$.
Define $\bm{\mathrm{\delta}}_t = (\delta_{t1}, \delta_{t2}, \ldots, \delta_{tM})^T$ and $\bm{\mathrm{\delta}}_r = (\delta_{r1}, \delta_{r2}, \ldots, \delta_{rN})^T$.
If the transmitter $m$ is in $\bm{\Gamma}_T$, then $\delta_{tm} = 1$; otherwise, $\delta_{tm}=0$. Similarly, if the receiver $n$ is in $\bm{\Gamma}_R$ then $\delta_{rn}=1$; otherwise,  $\delta_{rn}=0$. 

Consider $K$ moving targets in the \textit{x-y} plane.
Let $(x^k_{t}, y^k_{t})$ represent the location of the target $k=1,2,\ldots,K$ at time $t$.
As the targets move, the range and scattering angles of the channels change over time, and the time-varying SINR of the signal propagation channels are non-stationary with unknown expected rewards.

\subsection{Target dynamic modeling}\label{subsec:II-B}
For target $k=1, \ldots, K$, the target motion model is described by
\begin{equation}\label{eq1}
\bm{\mathrm{X}}^k_{t+1}=\bm{\mathrm{F}}^k_t(\bm{\mathrm{X}}^k_t)+\bm{\mathrm{G}}\bm{\mathrm{W}}_t^k,
\end{equation}
where ${\bm{\mathrm{F}}^k_t}(\cdot)$ denotes the dynamic state transition function of target $k$, which can be linear or non-linear and is defined according to the target dynamic mode. $\bm{\mathrm{X}}^k_t=[x^k_t, \dot{x}^k_t, y^k_t, \dot{y}^k_t]'$, where $[x^k_t, y^k_t]$ represents the position of target $k$ at time $t$, $[\dot{x}^k_t, \dot{y}^k_t]$ is the velocity of target $k$ at time $t$, and $(\cdot)'$ is the transpose operator.
The process noise $\bm{\mathrm{W}}_t^k=[\bm{\mathrm{W}}_{t,x}^k,\bm{\mathrm{W}}_{t,y}^k]'$, where $\bm{\mathrm{W}}_{t,x}^k=\bm{\mathrm{W}}_{t,y}^k
\sim N(0,\bm{\mathrm{Q}}_t^k)$ with covariance matrix being $\bm{\mathrm{Q}}_t^k=\text{diag}(Q_s,Q_s)$, where $Q_s$ is the process noise coefficient, and
\begin{equation}\label{eq3}
\bm{\mathrm{G}}=
\begin{bmatrix}
\frac{T_s^2}{2} & T_s & 0 & 0 \\
0 & 0 & \frac{T_s^2}{2} & T_s \\
\end{bmatrix}',
\end{equation}
where $T_s$ is the time interval.

\subsection{SINR modeling}\label{subsec:II-C}
A set of orthogonal waveforms in the MIMO radar system are transmitted, and each with a lowpass equivalent $s_m(t)$~\cite{godrich2011sensor}, where $\int_{{\tau}_{m}} |s_m(t)|^2=1$, and ${\tau}_{m}$ is the duration of the $m$-th transmitted signal. The waveform effective bandwidth is denoted by $\beta_m$. The transmitted power of the $m$-th signal is set to be $P_m$. Then the baseband representation for the signal $r_{mn}^k(t)$ from transmitter $m$ to receiver $n$ through target $k$ is 
\begin{equation}\label{eq4}
\begin{aligned}
r_{mn}^k(t)=&\sqrt{L_{t,mn}^k P_m}\psi_{t,mn}^k s_m\left(t-{\tau}_{mn}^k(t)\right)\\
& + \omega_{mn,\mathrm{noi}}(t) + \omega_{mn,\mathrm{int}_n}(t),
\end{aligned}
\end{equation}
where $L_{t,mn}^k\propto (R_{tm,t}^kR_{rn,t}^k)^{-2}$ represents the variation in the
signal strength due to path loss effects at time $t$. Let $R_{tm,t}^k=\sqrt{(x_{m_{Tx}}-x_{t}^k)^2 + (y_{m_{Tx}}-y_{t}^k)^2}$ and $R_{rn,t}^k=\sqrt{(x_{n_{Rx}}-x_{t}^k)^2 + (y_{n_{Rx}}-y_{t}^k)^2}$ represent the distances from the position $(x_{m_{Tx}}, y_{m_{Tx}})$ of transmitter $m$ to the position $(x_t^k, y_t^k)$ of target $k$ and from the location $(x_t^k, y_t^k)$ of target $k$ to the location $(x_{n_{Rx}}, y_{n_{Rx}})$ of receiver $n$ at time $t$. Let ${\tau}_{mn}^k(t)=(R_{tm,t}^k+R_{rn,t}^k)/{c}$ denote the propagation time from transmitter $m$ and receiver $n$ through target $k$ at time $t$ and $c$ is the speed of signal propagation. Define $\psi_{t,mn}^k$ as the radar cross section (RCS) of target $k$ impact on the phase and amplitude in channel \textit{m-n} at time $t$.  The noise $\omega_{mn,\mathrm{noi}}(t)$ and interference $\omega_{mn,\mathrm{int}_n}(t)$ at the receiver $n$ follow zero-mean, complex Gaussian noise distribution with autocorrelation function $\sigma_{\mathrm{noi}}^2\delta(\tau)$ and $\sigma_{\mathrm{int}_n}^2\delta(\tau)$, respectively.

The SINR metric is usually used to characterize the quality of the received signal, which is in general non-stationary for each TX-RX pair. Similar to~\cite{godrich2011sensor,pulkkinen2020reinforcement}, we omit the constants in the radar equation and obtain the SINR formula from transmitter $m$ to receiver $n$ for target $k$ at time $t$
\begin{equation}\label{eq5}
S_{mn}^k(t)={\bar{\mathrm{L}}_{mn}(\bm{\mathrm{X}}_t^k){\psi}_{t,mn}^k P_m}/({\sigma_{\mathrm{noi}}^2+\sigma_{\mathrm{int}_n}^2}),
\end{equation}
where $\bar{\mathrm{L}}_{mn}(\bm{\mathrm{X}}_t^k)=(R_{tm,t}^kR_{rn,t}^k)^{-2}$, and $\psi_{t,mn}^k=\psi_{t,mn}^{t,k} \psi_{t,mn}^{r,k}$ is a product of two scattering coefficients 
that are respectively taken from the monostatic target RCS model at the illumination angle and the scattering angle. That is, SINR is dependent on spatial distribution geometry.

\subsection{Measurement modeling}\label{subsec:II-D}
The measurement received by the distributed MIMO radar consists of the range $r$ and the azimuth $\phi$. 
The measurement of the target $k$ obtained via the signal propagation channel \textit{m-n} at time $t$ is
\begin{equation}\label{eq6}
\bm{\mathrm{Z}}_{mn}^k(t)=\mathrm{h}_{mn}(\bm{\mathrm{X}}_{t}^k)+\bm{\mathrm{\eta}}_{t,mn}^k,
\end{equation}
where the measurement function $\mathrm{h}_{mn}(\bm{\mathrm{X}}_{t}^k) = (r_{t,mn}(\bm{\mathrm{X}}_{t}^k),  \phi_{t,mn}(\bm{\mathrm{X}}_{t}^k))^T$ with $r_{t,mn}(\bm{\mathrm{X}}_{t}^k)=R_{tm,t}^k+R_{rn,t}^k$, $\phi_{t,mn}(\bm{\mathrm{X}}_{t}^k )=\text{arctan}((y_{t}^k-y_{n_{Rx}})/({x_{t}^k-x_{n_{Rx}}}))$.

Consider the measurement error $\bm{\eta}_{t,mn}^k \sim N ( 0, \bm{\mathrm{\sigma}}_{t,mn}^k )$ with covariance $\bm{\mathrm{\sigma}}_{t,mn}^k=\text{diag}(\sigma_{r,k,mn}^2(t),\sigma_{\phi,k,mn}^2(t))$, where $\sigma_{r,k,mn}^2(t)$ and $\sigma_{\phi,k,mn}^2(t)$ are the variances of the noise,
which are the CRLBs based on the maximum likelihood estimates of range and azimuth angle, respectively~\cite{dai2022adaptive,greco2011cramer,zhang2021efficient}. Define
\begin{equation}\label{eq9}
\begin{cases}
\begin{aligned}
\sigma_{r,k,mn}^2(t) \propto & (P_m\psi_{t,mn}^k \bar{\mathrm{L}}_{mn}(\bm{\mathrm{X}}_t^k)\beta_m^2)^{-1} \propto (S_{mn}^k(t)\beta_m^2)^{-1},
\end{aligned}
\\
\begin{aligned}
\sigma_{\phi,k,mn}^2(t) \propto & (P_m\psi_{t,mn}^k \bar{\mathrm{L}}_{mn}(\bm{\mathrm{X}}_t^k)\varsigma_n^2)^{-1} \propto (S_{mn}^k(t)\varsigma_n^2 )^{-1},
\end{aligned}
\end{cases}
\end{equation}
where $\varsigma_n$ denotes the 3dB receive beamwidth of the receiver $n$. In a radar system, the measurement precision of both the slant range and the azimuth depends on the instantaneous SINR. 
Without loss of generality, we assume that $\beta_i = \beta_j$ and $\varsigma_i = \varsigma_j$, $i \neq j$. 
In this context, 
the measurement noise covariance $\bm{\mathrm{\sigma}}_{t,mn}^k$ varies with the SINR of different channels, which are related to the geometrical relationship between target $k$ and the selected transmitter $m$ and receiver $n$.

Here, we focus on the TX-RX selection policy in the distributed MIMO radar system, where the correlation coefficient $\lambda_{t,mn}^k$ is defined based on the mean of the true SINR of the channel \textit{m-n} for target $k$ at time $t$. That is,
\begin{equation}\label{eq32}
\lambda_{t,mn}^k=\frac{S_{mn}^k(t)-S_{\min}}{S_{\max}-S_{\min}}+0.2,
\end{equation}
where the SINR set $\bm{S}_k^*=\lbrace S_{mn}^k(t):m=1,2,\ldots,M;\,n=1,2,\ldots, N;\, t=1,2,\ldots,T\rbrace$. $S_{\min}$ is the minimum value in $\bm{S}_k^*$. 
The maximum value $S_{\max}$ in $\bm{S}_k^*$ is set to be $10$, of which the SINR is individually equal to 10 dB and $\lambda=1$.
Hence, $\bm{\mathrm{\sigma}}_{t,mn}^k={\bm{\mathrm{\sigma}}_0}/{\lambda_{t,mn}^k}$, where $\bm{\mathrm{\sigma}}_0=\text{blkdiag}(\sigma_{r}^2(0), \sigma_{\phi}^2(0))$.
The matrix $\bm{\mathrm{\sigma}}_{t,mn}^k$ of the channel \textit{m-n} is potentially different for different target $k$ and time $t$ with respect to SINR. This scheme defined in \eqref{eq32} can distinguish each channel from the measurement error covariance matrix to reflect the scheduling performance in RMSE metric obviously and verify the correctness of the proposed MG-CRB-CL algorithm.

In a distributed MIMO radar system, the measurement vector $\bm{\mathrm{Z}}(t)$ at time $t$ consists of $M_s \cdot N_s \cdot K$ measurements for different targets and channels of the TX-RX pairs. That is, 
\begin{equation}\label{eq10}
\bm{\mathrm{Z}}(t+1)=[\ldots,\bm{\mathrm{Z}}^k(t+1)',\ldots]',
\end{equation}
where $\bm{\mathrm{Z}}^k(t+1)=[\ldots,\bm{\mathrm{Z}}_{mn}^k(t+1)',\ldots]'$, $\delta_{tm}=1$, and $\delta_{rn}=1$.

We further consider an instantaneous SINR measurement $\gamma_{mn}^k(t)$ for target $k$ that satisfies $\mathrm{E}[\gamma_{mn}^k(t)]=S_{mn}^k(t)$, where $\mathrm{E}[\cdot]$ represents the expectation operation. Similar to \cite{aittomaki2010performance}, we consider a case where the RCS of a target follows the Swerling-I model.
The SINR measurements of each channel \textit{m-n} can be derived by $\gamma_{mn}^k(t)\sim\text{Exp} (S_{mn}^k(t))$, 
which is exponentially distributed with the unknown mean $S_{mn}^k(t)$.

\subsection{MG-CRB model}\label{subsec:II-E}
For TX-RX pairs selection of the original problem, we need to associate a subset of the TX-RX pairs to a bandit process as a candidate selection of TX-RX pairs for our problem.
In this context, there are in total $C_M^{M_s} C_N^{N_S}$ different bandit processes when we apply a classic MAB framework.
Given a large number of transmitters and receivers in a distributed MIMO radar system in general, the classic MAB framework is usually computationally expensive due to enormous $C_M^{M_s} C_N^{N_S}$.
Meanwhile, the performance of the UCB1 algorithm is excellent on bandits with a small number of arms, but may degrade rapidly as the number of arms increases~\cite{kuleshov2014algorithms}.

Leveraging on the idea of CMAB technique,
we treat each TX-RX pair channel \textit{m-n} as a base arm. Therefore, the number of arms is $M \cdot N$, which is far less than $C_M^{M_S} C_N^{N_S}$. 
Hence, at each time, a super arm is played and the base arms contained in the super arm are played to observe their rewards.  

Different from the non-restless MAB model~\cite{chen2016combinatorial}, where the state of the arm keeps unchanged if not played, the RMAB model updates one arm’s state even though it is not played. 
Obviously, when tracking a moving target, the state of channels, i.e., the SINR, is time-varying with the geometry between the target and the transmitters and receivers.
So the sample mean of reward observation over all times cannot represent the approximated rewards of the arms, and the RMAB model considering the different combinations of  transmitter and receiver, namely, 
combinatorial restless bandit~(CRB),
is more appropriate in the case with moving targets. 

\begin{figure}[!t]
	\centering
	\includegraphics[width=2.2in]{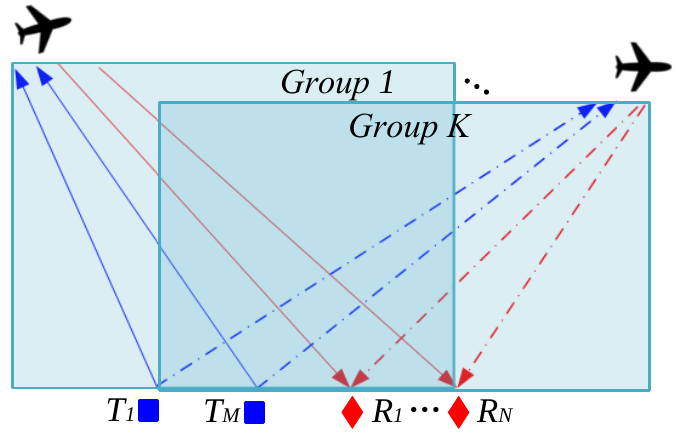}
	\caption{The MG-CRB model.}
    \label{fig:multigroup}
\end{figure}

Since the locations of targets are different most of the time, the geometric distribution between the MIMO radar system and target $k_1$ is different from the corresponding geometric distribution of target $k_2$, if $k_1\neq k_2$ and $k_1, k_2 \in \{1,\ldots, K\}$. 
The CRB model has to be expanded to a multi-group CRB (MG-CRB) architecture, which consists of $K$ groups and each group is associated with a CRB process representing a channel combination of a single target 
and computes corresponding SINR state $\hat{Y}_{mn}^k\left(t\right), m=1, \dots, M; n=1, \dots, N; k=1, \dots, K$. The MG-CRB model is established as shown in Fig.~\ref{fig:multigroup}, where the multiple solid lines and dotted lines denote the channels of two different groups. 
We fuse multi-group channels SINR state $\hat{Y}_{mn}^k\left(t\right)$ according to the weight $\Omega_k$, and establish the single objective optimization problem model. The optimal super arm can be selected to achieve the best multi-target tracking SINR level. 

In general, the performance of a policy is measured with regret, which quantifies the missed value from the cumulative reward in the non-stationary reward circumstance. In this paper, the policy is defined that the deterministic mapping from states to actions and
the performance for policy $\pi$ is measured with regret $\mathrm{R}_{\pi}(T)$, which is the difference between the cumulative reward obtained by always choosing the super arm $\bm{\mathcal{A}}_t^*=\lbrace\bm{\Gamma}_{T}^*,\bm{\Gamma}_{R}^* \rbrace$ with the highest known mean and the cumulative reward obtained with the chosen super arms of policy $\pi$,
\begin{equation}\label{eq13}
\mathrm{R}_{\pi}(T)=\sum_{t=1}^{T}\left[\mu_t^*-\mathrm{E}\left(\mu_t^{\pi}\right)\right],
\end{equation}
where $T$ is the time horizon, 
$\mu_t^{\pi}$ is the expected reward for the policy $\pi$ at time $t$ and $\mu_t^*=\mathop{\max}\limits_{\bm{\mathcal{A}}} \sum_{(m,n) \in\bm{\mathcal{A}}} \sum_{k=1}^{K} \Omega_k S_{mn}^k(t)$ is the expected reward for the optimal super arm $\bm{\mathcal{A}}_t^*$ at time $t$. The right-hand side $\sum_{t=1}^{T}\mathrm{E}\left(\mu_t^{\pi}\right)$ denotes the total expected reward of policy $\pi$ over the time horizon $T$.

Intuitively, we would like the regret $\mathrm{R}_{\pi}\left(T\right)$ to be as small as possible. If it is sub-linear with respect to time horizon $T$, 
the time-averaged regret will tend to zero.

\subsection{Problem statement}\label{subsec:II-F}
Considering the distributed MIMO radar system in Fig.~\ref{fig:multigroup}, our purpose in this paper is to jointly solve the multi-target tracking problem and the TX-RX selection problem simultaneously.
The former aims to 
estimate the states of the multiple targets
in the surveillance area given the  measurements $\bm{\mathrm{Z}}(t+1)$ received by the selected channels via the Bayes rule;
while the latter aims to find the best channels to observe the targets given the available information up to the decision time.  
In the TX-RX selection, we regard the multi-target SINR performance as the system reward. 
Note that $S_{mn}^k(t)$ for the target $k$ at time $t$ is not known when the channel \textit{m-n} is not probed. 
Since, in general, it is impossible to probe all the channels at the same time, there is a trade-off between exploring different channel subsets and exploiting the best channel known so far.
In this paper, we aim to improve SINR of a distributed MIMO radar system in the multi-target tracking process and maximize the cumulative long-run reward over the time horizon $T$, 
that is, $\sum_{t=1}^{T}\mathrm{E}\left(\mu_t\right)$~(See \eqref{eq13}). 
The  maximization problem for multi-target tracking can be formulated as
\begin{equation}\label{eq17}
\begin{aligned}
&\max\limits_{\bm{\delta}_t^{1:T},\bm{\delta}_r^{1:T}}~\mathrm{R}\left(\bm{\delta}_t^{1:T},\bm{\delta}_r^{1:T}\right) \\
=&\max\limits_{\bm{\delta}_t^{1:T},\bm{\delta}_r^{1:T}} \mathrm{E} \left\{ \sum_{t=1}^{T} \left[\sum_{k=1}^{K}\sum_{m=1}^{M}\sum_{n=1}^{N} \delta_{tm}^{t}\delta_{rn}^{t}\Omega_k\gamma_{mn}^k(t)\right] \right\}, \\
&\bm{\text{s.t.}}~~~
\text{sum}\left(\bm{\delta}_t^t\right)=M_s,~
\text{sum}\left(\bm{\delta}_r^t\right)=N_s.
\end{aligned}
\end{equation}
where $\bm{\delta}_{t}^{1:T}=\{\bm{\delta}_{t}^1,\ldots,\bm{\delta}_{t}^T\}$ and $\bm{\delta}_{r}^{1:T}=\{\bm{\delta}_{r}^1,\ldots,\bm{\delta}_{r}^T\}$. $\bm{\delta}_{t}^t (\bm{\delta}_{r}^t)$ is the transmitters (receivers) selection vectors at time $t$, 
$\gamma_{mn}^k(t)$ is the instantaneous SINR measurement from target $k$ at time $t$, and $\Omega_k$ denotes the weight of target $k$, $k=1,\ldots,K$. 

\section{MG-CRB-based closed-loop optimization framework}\label{sec:solutions}
\begin{figure}[!t]
\centering
\includegraphics[width=3.5in]{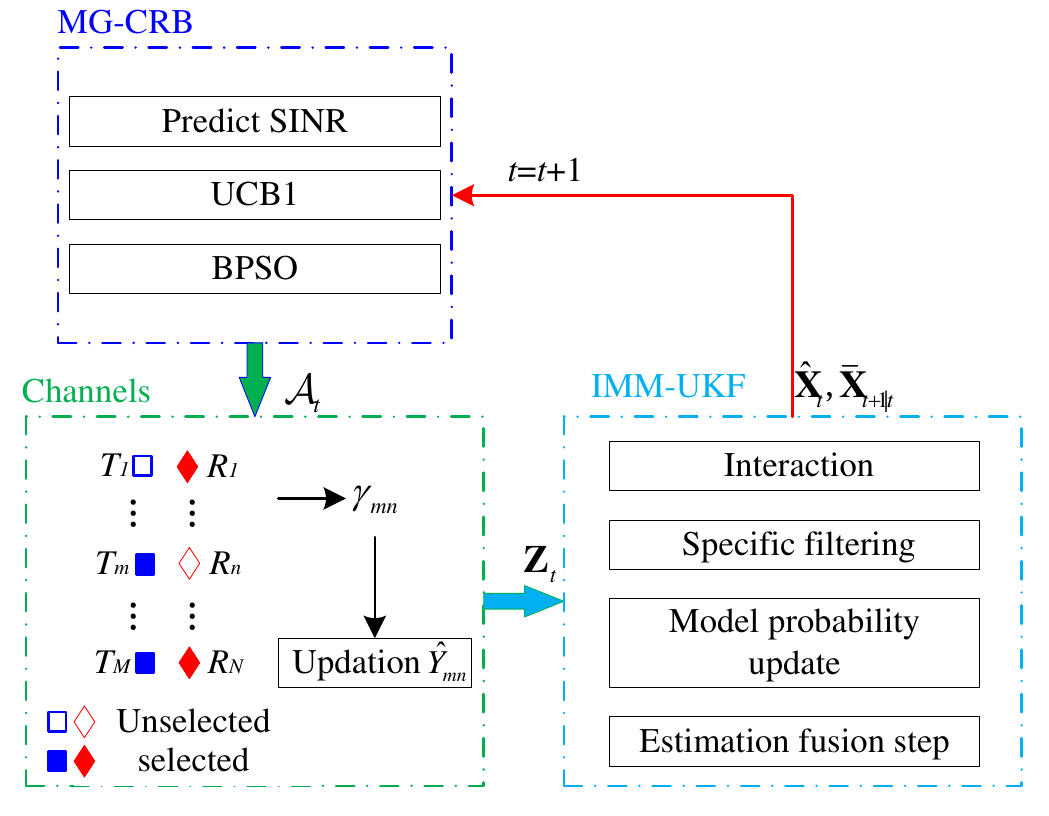}
\caption{The closed-loop optimization framework.}
\label{fig:framework}
\end{figure}

In this section, we develop the arm selection scheme based on UCB1 in MG-CRB model and adapt BPSO method to search the global optimal super arm, which consists of $M_s \cdot N_s$ channels for a distributed MIMO radars system. IMM-UKF method is adapted to estimate the target state at the current time and predict the target state at the next time. 
Then, the closed-loop framework is established to achieve the joint optimization between the filtering process and the channel selection, which is shown in Fig.~\ref{fig:framework}.

\subsection{The arm selection scheme in MG-CRB model}\label{subsec:III-A}
We introduce the UCB1 algorithm~\cite{chen2016combinatorial} and the counters $\hat{Y}_{mn}^k\left(t\right)$ and $\mathrm{d}_{mn}^k\left(t\right)$ that track the sample mean of the rewards obtained from arm \textit{m-n} up to the current time slot, and the number of times the arm has been played in group $k$, respectively, which are updated at time $t$ as
\begin{equation}\label{eq11}
\hat{Y}_{mn}^k\left(t\right)=\begin{cases}
\dfrac{\hat{Y}_{1:t-1,mn}^k+\gamma_{mn}^k\left(t\right)}{\mathrm{d}_{mn}^k\left(t-1\right)+1},
& \text{if played} \\
\hat{Y}_{mn}^k\left(t-1\right), & \text{otherwise.} 
\end{cases}
\end{equation}
\begin{equation}\label{eq12}
\mathrm{d}_{mn}^k\left(t\right)=\begin{cases}
\mathrm{d}_{mn}^k\left(t-1\right)+1, & \text{if played}\\
\mathrm{d}_{mn}^k\left(t-1\right), & \text{otherwise.}
\end{cases}
\end{equation}
where $\hat{Y}_{1:t-1,mn}^k\triangleq\mathrm{d}_{mn}^k\left(t-1\right)\hat{Y}_{mn}^k\left(t-1\right)$ denotes the sample sum reward up to time $t-1$,
and $\gamma_{mn}^k\left(t\right)$ is the reward observation of arm \textit{m-n} for target $k$ at time $t$.

In the MG-CRB model established in Section~\ref{subsec:II-E}, we use the target dynamic state to provide a prior SINR information for the next time, not just the sample value in the group $k$.
If the arm \textit{m-n} is played, the state for estimating the expected SINR reward can be fused by $\gamma_{mn}^k\left(t\right)$ and the predicted SINR mean reward $\bar{Y}_{mn}^k\left(t\right)$ with constant coefficient $\bar{\alpha}$. Otherwise, the state is equal to $\bar{Y}_{mn}^k\left(t\right)$, which can be predicted by the filtered dynamic state $\hat{\bm{\mathrm{X}}}_{t-1}^k$ and the predicted target dynamic state $\bar{\bm{\mathrm{X}}}_{t}^k$ of the target - both are derived from IMM-UKF described in Section \ref{subsec:III-C}. 
In other words,
\begin{equation}\label{eq14}
\hat{Y}_{mn}^k\left(t\right)=\begin{cases}
\left(1-\bar{\alpha}\right)\bar{Y}_{mn}^k\left(t\right)+\bar{\alpha}\gamma_{mn}^k\left(t\right), & \text{if played}\\
\bar{Y}_{mn}\left(t\right), & \text{otherwise.}
\end{cases}
\end{equation}
\begin{equation}\label{eq15}
\mathrm{d}_{mn}^k\left(t\right)=\begin{cases}
\mathrm{d}_{mn}^k\left(t-1\right)+1, & \text{if played}\\
\mathrm{d}_{mn}^k\left(t-1\right), & \text{otherwise.}
\end{cases}
\end{equation}
where $\bar{Y}_{mn}^k\left(t\right)=\mathrm{\Psi}(\hat{Y}_{mn}^k\left(t-1\right),\bar{\bm{\mathrm{X}}}_t^k,\hat{\bm{\mathrm{X}}}_{t-1}^k)$ is the predicted SINR mean reward, and $\bar{\alpha} \in \left(0,1\right)$ is a constant.

We consider constant noise power $\sigma_{\mathrm{noi}}^2$ and the inference power $\sigma_{\mathrm{int}_n}^2$ of a channel over time horizon $T$.
Our method can be directly extended to the case with time-varying noise and interference power provided that their dynamic models are available.
The values of the noise and interference power are unknown to the central controller of the distributed MIMO radar system, and $\bar{Y}_{mn}^k(t)$ cannot be directly derived by $\bar{\bm{\mathrm{X}}}_t^k$. 
Hence, based on \eqref{eq5}, we can adapt the ratio between the SINR approximation at time $t$ and $t-1$ by using the target kinematic states $\bar{\bm{\mathrm{X}}}_t^k$ and $\hat{\bm{\mathrm{X}}}_{t-1}^k$,
\begin{equation}\label{eq16}
\begin{aligned}
& \Psi(\hat{Y}_{mn}^k(t-1), \bar{\bm{\mathrm{X}}}_t^k,\hat{\bm{\mathrm{X}}}_{t-1}^k) \\ =&\alpha\frac{\bar{\mathrm{L}}_{mn}(\bar{\bm{\mathrm{X}}}_t^k)\psi_{t,mn}^k\hat{Y}_{mn}^k\left(t-1\right)}{\bar{\mathrm{L}}_{mn}(\hat{\bm{\mathrm{X}}}_{t-1}^k)\psi_{t-1,mn}^k} + \left(1-\alpha\right)\hat{Y}_{mn}^k\left(t-1\right),
\end{aligned}    
\end{equation}
where $\alpha \in \left(0,1\right)$ is a constant. 
$\bar{\mathrm{L}}_{mn}(\cdot)$ and $\psi_{t,mn}^k$ are relevant functions, which can be computed through \eqref{eq5} with given $\bar{\bm{\mathrm{X}}}_t^k$ and $\hat{\bm{\mathrm{X}}}_{t-1}^k$ and channel estimation~\cite{shen2013power,li2013robust}, respectively, where $m=1,2, \dots, M$, $n=1,2, \dots, N$.

Based on the prediction in \eqref{eq16} on the SINR state of each arm from the last time slot, we can establish the index value of each arm of each group at time $t$, in the vein of the UCB1 algorithm~\cite{auer2002finite}.
\begin{equation}\label{eq18}
\mathrm{I}_{mn}^k\left(t\right)=\bar{Y}_{mn}^k\left(t\right)+\sqrt{{\beta\ln{t}}/{\mathrm{d}_{mn}^k(t)}},
\end{equation}
where $\beta$ is a positive constant coefficient.
Then we select the super arm to complete the TX-RX pair subset selection according to the weight $\Omega_k$ of target $k$. The fused index value of each arm is given by
\begin{equation}\label{eq34}
\mathrm{I}_{mn}\left(t\right)=\sum_{k=1}^K \Omega_k \mathrm{I}_{mn}^k\left(t\right),
\end{equation}

The super arm $\bm{\mathcal{A}}_t$ is selected that maximizes $\sum_{\left(m,n\right)\in\bm{\mathcal{A}}_t}\mathrm{I}_{mn}\left(t\right)$.

It is worth noting that, compared with the original problem of selecting radar stations, the channel selection problem needs an additional constraint that selects channels that correspond to different transmitters and receivers, 
and hereby can obtain the same optimum as the original problem. It is straightforward to select $M_s \cdot N_s$ arms under the constraint that the corresponding channels come from $M_s$ different transmitters and $N_s$ different receivers simultaneously, instead of selecting the $M_s \cdot N_s$ arms that can take higher SINR value pay-off. For example, in a MIMO radar system with $M_s=2,N_s=2$, the correct selected channels set is assumed as $\{$\text{1-1}, \text{1-2}, \text{2-1}, \text{2-2}$\}$, instead of $\{$\text{1-1}, \text{1-2}, \text{1-3}, \text{2-2}$\}$, whose signal is received by $3\neq N_s$ stations. 
Hence, the combination scheme in the MG-CRB-CL algorithm, 
which will be presented in the following Section~\ref{subsec:III-B},
is used to select arms to form a super arm for the MIMO radar system.

\subsection{BPSO for super arm selection}\label{subsec:III-B}
Since the constraint that the selected corresponding channels come from $M_s$ different transmitters and $N_s$ different receivers simultaneously, instead of the $M_s \cdot N_s$ arms that can take higher value pay-off, whose example is described in Section \ref{subsec:III-A}, combinatorial optimization for obtaining the best super arm at each time also matters.
In order to avoid the computational burden and exponential explosion of the exhaustive search method, we adapt the BPSO method~\cite{wu2022optimal} to achieve the heuristic search for global optimal solutions.
The computational complexity of BPSO mainly depends on the cardinality $N_{\text{pop}}$ of swarms and the maximum number of iterations $q_{\max}$. 

Learning from the foraging procedure of bird flocking, the PSO is a kind of population-based stochastic optimization algorithm, where each individual (particle) is represented for an alternative solution. All the particles simultaneously move in the solution space of the optimized objective function to search for the optimal solution.
Holding attributes, such as position, velocity, and fitness, all particles will update them by tracking their own best-known particle and the best swarm up to the current iteration. 
The best particle and swarm are evaluated and sorted by fitness value, which can be set through objective functions. 
Through multiple iterations and particle swarms cooperation, the global optimal solution can be obtained. 
Based on PSO, BPSO defines the position of each particle with a certain amount of binary ``0'' and ``1'' numbers.
The velocity is defined as probability, which is computed by the Sigmoid function, and represents the mutation possibility that position may change to a different number ``0'' or ``1''~\cite{el2013binary}.

Considering the constraint in \eqref{eq17} and reducing the number of position bits as many as possible, we combine the transmitters selection vector $\bm{\delta}_{t}$ and receivers selection vector $\bm{\delta}_{r}$ to design the position vector of each particle as $\bm{\mathrm{D}}=\left[\bm{\delta}_{t},\bm{\delta}_{r}\right]$.
The length of the position vector $\bm{\mathrm{D}}$ of particles is equal to $M+N$ and the number ``1'' in position bits represents that the corresponding transmitter or the receiver is selected in this solution; otherwise, the corresponding transmitter or the receiver is not selected.

The fitness of particles is evaluated by the current index value $\mathrm{I}_{mn}(t)$ of corresponding channels, which come from the $M_s$ transmitters and $N_s$ receivers represented by position vectors of particles.
Note that the fitness is zero when the position of one particle dissatisfies the constraint in \eqref{eq17}. 
The fitness function at time $t$ for particle $i$ of the $q$-th iteration is
\begin{equation}\label{eq19}
F_q^i(t)=\begin{cases}
\sum\limits_{m=1}^{M}\sum\limits_{n=1}^{N} {\delta}_{tm}^{q,i} {\delta}_{rn}^{q,i} \mathrm{I}_{mn}\left(t\right),
 & \begin{aligned}
     \text{if sum} & (\bm{\delta}_t^{q,i})=M_s, \\ \text{sum}& (\bm{\delta}_r^{q,i})=N_s.
 \end{aligned}
  \\
0, & \text{otherwise.}
\end{cases}
\end{equation}
where $\bm{\delta}_{t}^{q,i} (\bm{\delta}_{r}^{q,i})$ is the transmitters (receivers) selection vector of particle $i$ in the $q$-th iteration, which is contained in the position vector $\bm{\mathrm{D}}_q^i$ of particle $i$ in the $q$-th iteration, $i=1, \dots, N_{\text{pop}}$ and $q=1, \dots, q_{\max}$.

Based on the fitness value of the particle swarm in each iteration, the best solution $\bm{\mathrm{D}}^*_i$ of particle $i$ up to iteration $q$ and the global best solution $\bm{\mathrm{D}}^*_g$ from the entire swarm until iteration $q$ is obtained. At the iteration step $q+1$, the velocity $\bm{\mathrm{V}}_{q+1}^i$ and position $\bm{\mathrm{D}}_{q+1}^i$ are calculated as 
\begin{equation}\label{eq20}
\bm{\mathrm{V}}_{q+1}^i=\omega_t \bm{\mathrm{V}}_{q}^i + c_1 r_1 ( \bm{\mathrm{D}}^*_i-\bm{\mathrm{D}}_q^i ) + c_2 r_2 ( \bm{\mathrm{D}}^*_g-\bm{\mathrm{D}}_q^i ),
\end{equation}
where $\omega_t$ is a weighting coefficient, which is decreased over the period of iterations. Parameters $c_1$ and $c_2$ are arbitrary constants, and $r_1$ and $r_2$ are random variables from the uniform distribution $\text{U}(0,1)$. 
\begin{equation}\label{eq21}
{D}_{q+1}^{i,j}=\begin{cases}
\text{XOR}({D}_{q}^{i,j},1 ), & \text{if}~\delta < \text{Sig}( {V}_{q+1}^{i,j}) \\
{D}_{q}^{i,j}, & \text{otherwise.}
\end{cases}
\end{equation}
where ${V}_{q+1}^{i,j}$ denotes the \textit{j}-th bit of $\bm{\mathrm{V}}_{q+1}^i$, ${D}_{q+1}^{i,j}$ denotes the $j$-th bit of $\bm{\mathrm{D}}_{q+1}^i$, $j=1, \ldots, M+N$. XOR is an xor operation with 1, 
and $\delta$ is random variable from the uniform distribution $\text{U}(0,1)$. $\text{Sig}(\cdot)$ denotes the Sigmod function.

Based on the above description, the BPSO method iteratively searches the global best solution $\bm{\mathrm{D}}^*_g$ until $q_{\max}$ is reached and selects the best super arm $\bm{\mathcal{A}}_t$ that holds the highest probability to obtain the maximum instantaneous reward at time $t$. 
After the super arm $\bm{\mathcal{A}}_t$ is played, the SINR state $\hat{Y}_{mn}^k(t)$ of each arm of the group $k$ at time $t$ is 
updated by $\gamma_{mn}^k(t)$ and the predictions $\bar{Y}_{mn}^k(t)$.

\subsection{IMM-UKF for target state estimation}\label{subsec:III-C}
When the super arm $\bm{\mathcal{A}}_t$ has been selected at time $t$, the radar measurement vector $\bm{\mathrm{Z}}_t$ can be obtained, which consists of $M_s\cdot N_s \cdot K$ measurements of multiple TX-RX pairs and multiple targets. Given that the data association processing has been completed, the tracking filtering processes of the targets can be solved separately and described as follows.

To manage the nonlinear measurement function varying across different channels, we employ a distributed track fusion approach and utilize the IMM-UKF method for the state estimation $\hat{\bm{\mathrm{X}}}_{t,mn}^k$ of target $k$ and channel \textit{m-n}.
The state of target $k$ is obtained by fusing the state estimations of all $M_s\cdot N_s$ channels in group $k$.

The initial state ${\bm{\mathrm{X}}}_{t=1,mn}^k$ and the initial covariance ${\bm{\mathrm{P}}}_{t=1,mn}^k$ of target $k$ in channel \textit{m-n} are defined firstly. 
$\hat{\bm{\mathrm{X}}}_{t=1,mn}^{k,i}$ and $\hat{\bm{\mathrm{P}}}_{t=1,mn}^{k,i}$ denote the target state estimation and the corresponding covariance matrix of the $i$-th model, respectively. 
The initial model probability vector is $\hat{\bm{\mathrm{u}}}_{t=1,mn}^k=[\hat{u}_{t=1,mn}^{k,1},\hat{u}_{t=1,mn}^{k,2},\dots,\hat{u}_{t=1,mn}^{k,M_{\text{model}}}]'$, where $M_{\text{model}}$ is the number of models. 
$\hat{\bm{\mathrm{X}}}_{t=1,mn}^{k,i}$ and $\hat{\bm{\mathrm{P}}}_{t=1,mn}^{k,i}$ are initialized as ${\bm{\mathrm{X}}}_{t=1,mn}^k$ and ${\bm{\mathrm{P}}}_{t=1,mn}^k$, $i=1,\dots,M_{\text{model}}$, respectively. 
The procedures of IMM-UKF consist of interaction,  filtering, model probability update, and estimation fusion steps.
For details, the reader is referred to \cite{dai2022adaptive,xu2017cost}.

After these procedures being carried out, the model probabilities $\hat{\bm{\mathrm{u}}}_{t,mn}^k$ and the combined state of multiple models $\hat{\bm{\mathrm{X}}}_{t,mn}^k$ are estimated at time $t$ from channel \textit{m-n}. 
Finally, the fused target state is
\begin{equation}\label{eq22}
\hat{\bm{\mathrm{u}}}_{t}^k=\frac{\sum_{(m,n)\in\bm{\mathcal{A}}_{t}}\hat{\bm{\mathrm{u}}}_{t,mn}^k}{M_sN_s},~~
\hat{\bm{\mathrm{X}}}_{t}^k=\frac{\sum_{(m,n)\in\bm{\mathcal{A}}_{t}}\hat{\bm{\mathrm{X}}}_{t,mn}^k}{M_sN_s},
\end{equation}
where $\hat{\bm{\mathrm{u}}}_{t}^k=[\hat{u}_{t}^{k,1},\ldots,\hat{u}_{t}^{k,M_{\text{model}}}]'$.

Then the prediction of the dynamic state at time $t+1$ for target $k$ is given by
\begin{equation}\label{eq24}
\bar{\bm{\mathrm{X}}}_{t+1}^k=\sum\nolimits_{i=1}^{M_{\text{model}}} \hat{u}_{t}^{k,i}  \bm{\mathrm{F}}_t^{k,i}(\hat{\bm{\mathrm{X}}}_t^k),
\end{equation}
where $\bm{\mathrm{F}}_t^{k,i}\left(\cdot\right)$ is the state transition function of the $i$-th model for target $k$ at time $t$, $i=1,\dots,M_{\text{model}}$.

\subsection{Closed-loop optimization framework}\label{subsec:III-D}
We aim to improve the SINR level of the radar system to strengthen the performance of target tracking. 
Therefore, a closed-loop optimization framework is established, where the main idea is that the target tracking information at the current time can be used for the prediction of the SINR state of each arm at the next time instead of only using the SINR sample value. 
Obviously, the more accurate the SINR state is predicted, 
the best channels are more likely to be selected with smaller errors for measurements. 
Also, the better performance of target tracking is obtained.

Unlike UCB1 in the special case of CMAB where all the inactive bandit processes are frozen, our MG-CRB-CL algorithm uses IMM-UKF to track the target states at the current time based on the measurement vector $\bm{\mathrm{Z}}_t$ and then uses the current target states to predict those at the next time slot.
Hence, the SINR state of each arm can be predicted or updated in the MG-CRB model, which plays an important part in the TX-RX selection policy and improves the SINR performance. 
We select the super arm $\bm{\mathcal{A}}_{t+1}$ and obtain the measurement vector $\bm{\mathrm{Z}}_{t+1}$ at the next time slot.
The procedures of the closed-loop tracking and selection framework are listed as follows.
\begin{itemize}
\item Step 1: Obtain the target state estimation $\hat{\bm{\mathrm{X}}}_{t}^k$ and prediction $\bar{\bm{\mathrm{X}}}_{t+1}^k$ at the current time $t$ from the measurement vector $\bm{\mathrm{Z}}_t$, which is derived by \eqref{eq22} and \eqref{eq24} with given $\bm{\mathcal{A}}_{t}$;

\item Step 2: For the arm \textit{m-n} of each group in the selected super arm $\bm{\mathcal{A}}_{t}$, the SINR state $\hat{Y}_{mn}^k(t)$ can be updated with a newly collected sample; otherwise, the predicted SINR value is saved as the $\hat{Y}_{mn}^k(t)$, $k=1,\ldots,K$ based on \eqref{eq14};

\item Step 3: Predict the SINR state $\bar{Y}_{mn}^k(t+1)$ of each channel of group $k$ at time $t+1$ by \eqref{eq16}, and select the best super arm $\bm{\mathcal{A}}_{t+1}$ based on the UCB1 index value $\mathrm{I}_{mn}\left(t+1\right)$ of SINR by \eqref{eq34};

\item Step 4: The measurement vector $\bm{\mathrm{Z}}_{t+1}$ at time $t+1$ is obtained based on the selection $\bm{\mathcal{A}}_{t+1}$.
\end{itemize}

Overall, for the non-stationary rewards, we use the SINR approximation with IMM-UKF and UCB1 index policy to effectively solve the TX-RX subset selection problem. 
The pseudo codes of our MG-CRB-CL algorithm are listed in Algorithm \ref{algorithm:Algorithm 1}.
\begin{algorithm}[!t]
\DontPrintSemicolon
\SetAlgoLined
\caption{The MG-CRB-CL algorithm}
\label{algorithm:Algorithm 1}
\tcp*[l]{\textbf{Initialization}}
\For{$k \gets 1$ \KwTo $K$}{
\For{$m \gets 1$ \KwTo $M$}{
   \For{$n \gets 1$ \KwTo $N$}{
    Play arm \textit{m-n} and initialize $(\hat{Y}_{mn}^k(0))_{M \times N}$, $(\mathrm{d}_{mn}^k(0))_{M \times N}$, respectively;\;
  }
  }
}
\tcp*[l]{\textbf{Main loop}}
\For{$t \gets 1$ \KwTo $T$}{
Use BPSO to select multiple arms to obtain the super arm $\bm{\mathcal{A}}_t$ that maximizes $\sum_{\left(m,n\right)\in\bm{\mathcal{A}}_t} \mathrm{I}_{mn}(t)$ and the target measurement vector $\bm{\mathrm{Z}}_t$;\;

\For{$k \gets 1$ \KwTo $K$}{
Update the target dynamic state $\hat{\bm{\mathrm{X}}}_{t}^k$;\;

Update $(\hat{Y}_{mn}^k(t))_{M \times N}$, $(\mathrm{d}_{mn}^k(t))_{M \times N}$ by \eqref{eq14} and \eqref{eq15};\;

Compute the predicted state $\bar{\bm{\mathrm{X}}}_{t+1}^k$;\;

Compute the predicted SINR $\bar{{Y}}_{mn}^k\left(t+1\right)$ at time $t+1$ by \eqref{eq16}.\; 
}
}
\end{algorithm}

\subsection{Computational complexity}\label{subsec:III-E}
In this subsection, we evaluate the computational complexity of MG-CRB-CL, which uses BPSO to solve the combinatorial optimization problem. 
Therefore, the computational complexity is mainly affected by two factors: the cardinality $N_{\text{pop}}$ of swarms and the maximum number of iterations $q_{\max}$. In a specific run of each time, it requires the complexity of $\mathrm{O}(N_{\text{pop}}q_{\max})$. These parameters can be tuned by the performance of the MG-CRB-CL algorithm. However, for solving the TX-RX pairs selection problem, the exhaustive search algorithm needs a combinatorial complexity of $\mathrm{O}(C_M^{M_S}C_N^{N_S})$. The computational complexity comparison shows that MG-CRB-CL achieves a lower computation cost and enables real-time resource scheduling for multi-target tracking.

\section{Simulation and Results}\label{sec:simulation}
In this section, we compare our proposed MG-CRB-CL algorithm with the UCB1 algorithm and the $\epsilon$-greedy algorithm via simulation. 
We consider a distributed MIMO radar with $M=4$ transmitters and $N=6$ receivers in a 100~km $\times $ 100~km region, 
It is assumed that the MIMO radar system with widely separated transmit/receive antenna subarrays emits the wide beam to observe multiple targets simultaneously and the signal processing is accomplished by digital transceiver beamforming and match filter with orthogonal signals.

\begin{figure*}[!t]
    \begin{minipage}[t]{0.33\linewidth}
        \centering
        \includegraphics[width=2.0in]{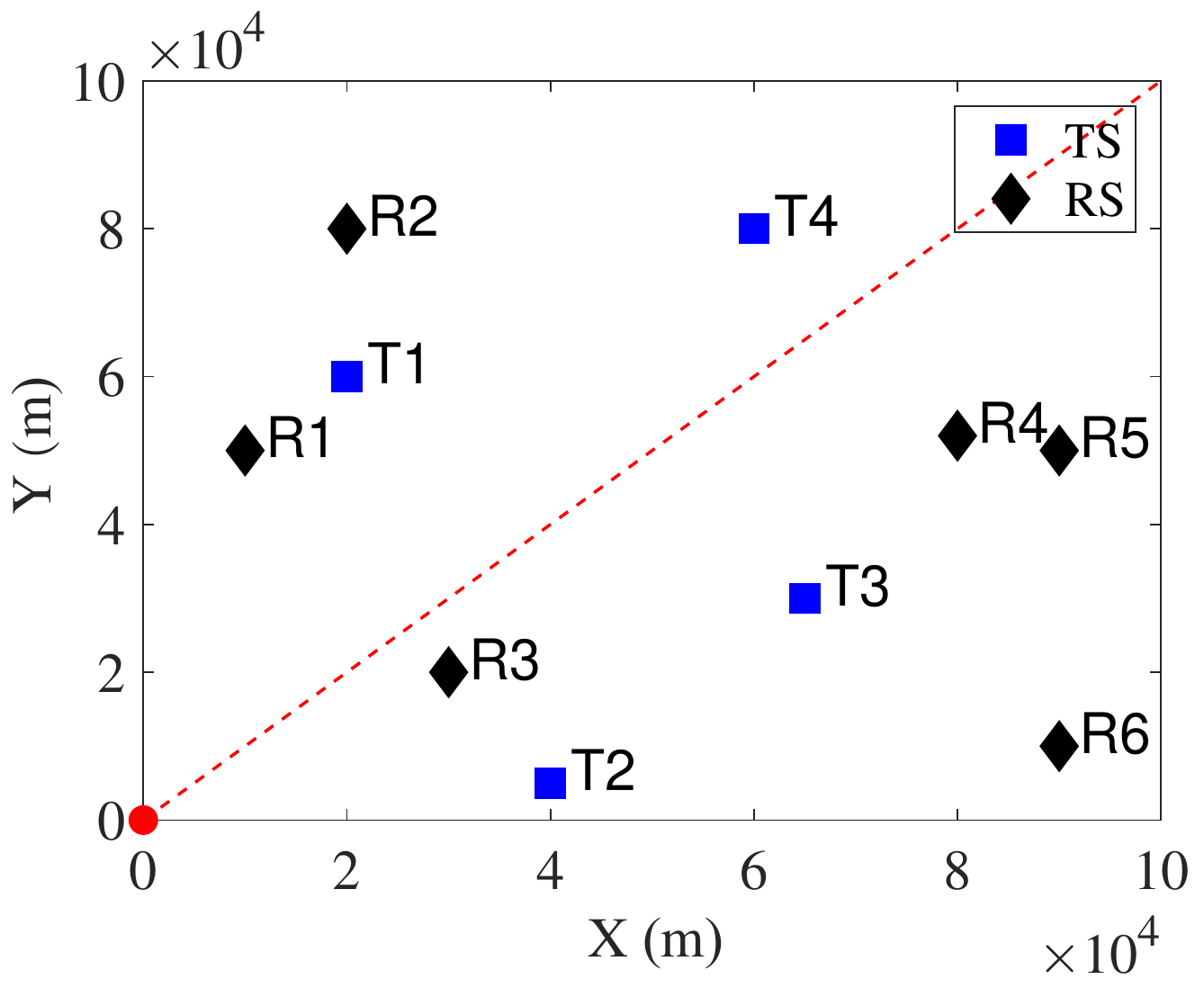}
        \centerline{(a)}
    \end{minipage}%
    \begin{minipage}[t]{0.33\linewidth}
        \centering
        \includegraphics[width=2.0in]{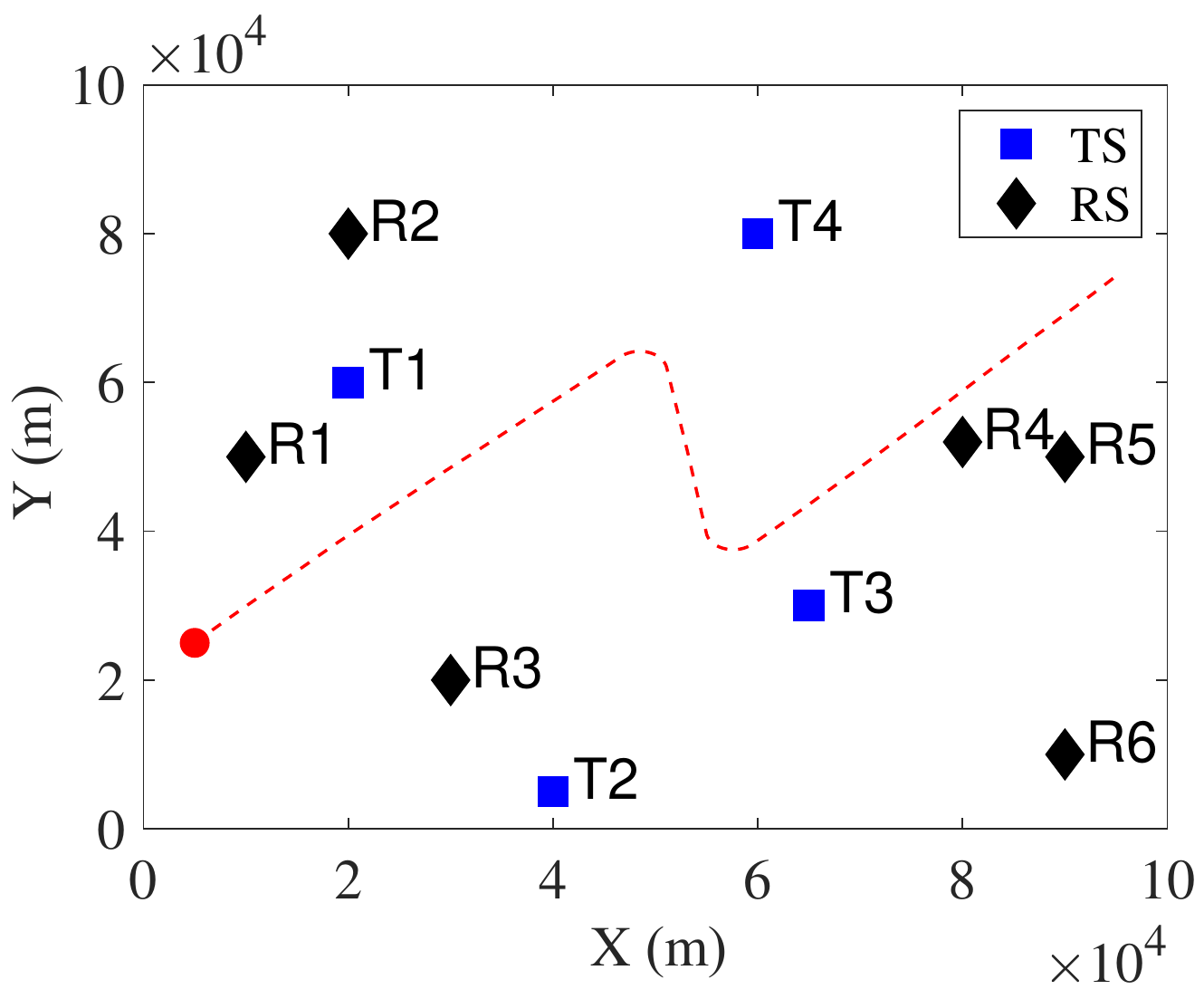}
        \centerline{(b)}
    \end{minipage}
    \begin{minipage}[t]{0.33\linewidth}
        \centering
        \includegraphics[width=2.0in]{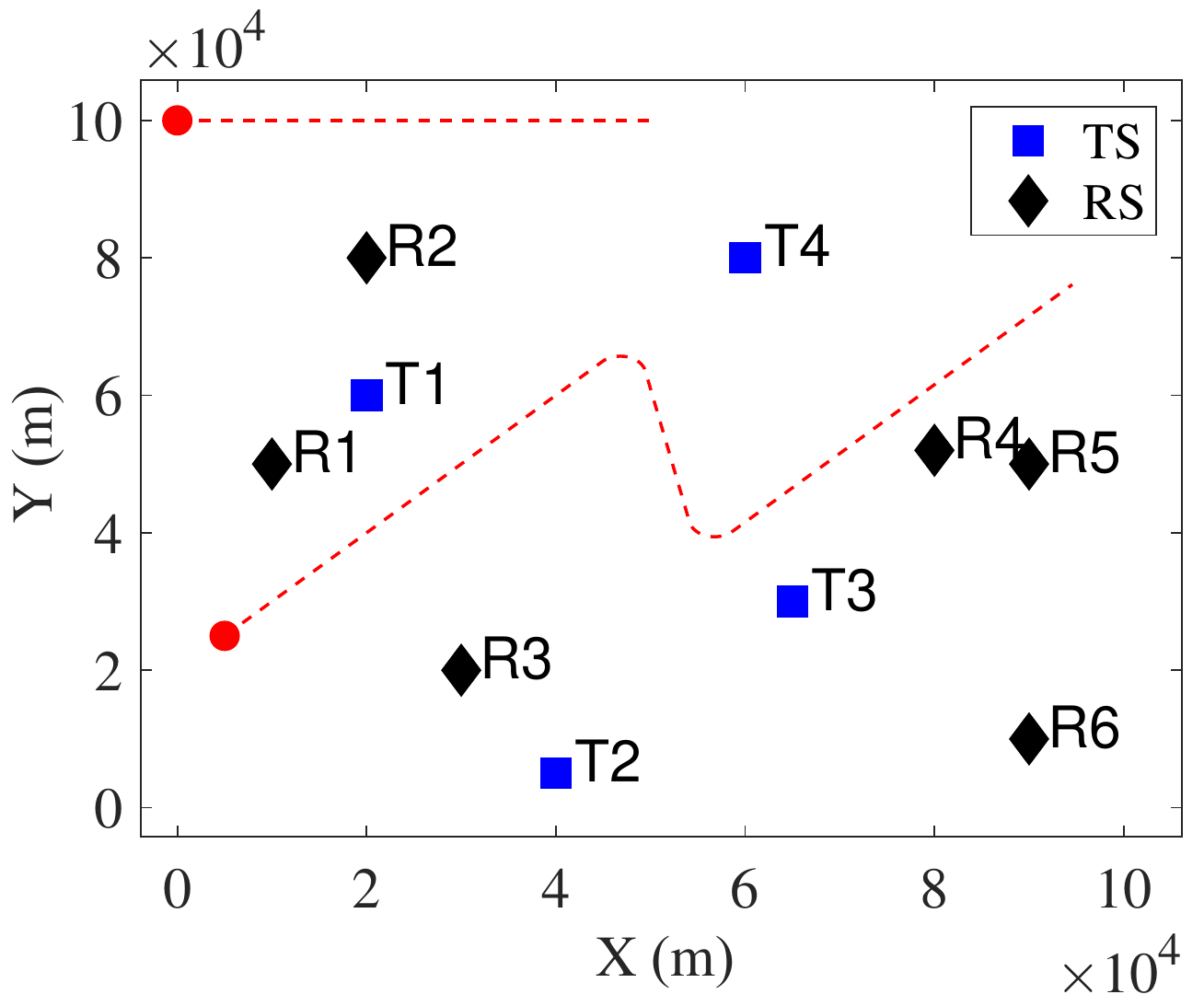}
        \centerline{(c)}
    \end{minipage}
    \caption{Settings of targets and all the radar stations in scenarios. (a) Scenario 1. (b) Scenario 2. (c) Scenario 3.}
    \label{fig_sim5}
\end{figure*}

Two single target tracking scenarios, which are different on target motion models, and a multiple targets tracking scenario are designed to show the effectiveness of MG-CRB-CL with $M_c=500$ Monte-Carlo trials, as shown in Fig.~\ref{fig_sim5}, where ``TS" represents a transmitter station, and ``RS" represents a receiver station. The starting points of targets are located in the red dots, and the red dotted lines represent the ground-truth tracks. For showing the effectiveness of the three selection algorithms, two fixed selection strategies are used and the corresponding super arms are $\bm{\mathcal{A}}_{\text{fix1}}=\lbrace[3,4],[1,3,4]\rbrace$ and $\bm{\mathcal{A}}_{\text{fix2}}=\lbrace[3,4],[4,5,6]\rbrace$. 
We use SINR cumulative regret and RMSE metrics to evaluate the system performances of all the compared algorithms.

We consider that the target reflectivity is uniform with respect to the channel \textit{m-n}, $\forall m,n$, which means that $\psi_{mn}=1~\mathrm{m}^2$, and the RCSs of the targets are 1 dB$\text{m}^2$.
For all the channels \textit{m-n}, $m=1, \dots, M$, $n=1, \dots, N$,
we set transmit power $P_m=1$ kW, signal effective bandwidth $\beta_m=0.1$ MHz and 3dB receive beamwidth $\varsigma_n=3^{\circ}$. 
The sampling time interval $T_s=1$ s. The thermal noise power $\sigma_{\mathrm{noi}}^2=10^{-26}$ and the interference power $\sigma_{\mathrm{int}}^2=\left[0.5,0.5,0.5,0.5,0.5,0.5\right]*10^{-21}$, which are constant over time.

Three models are used in IMM-UKF to track the targets, including one nearly constant velocity (NCV) model and two nearly constant turn rate (NCT) models with the angular speed of turning $\omega \in \{3^{\circ}, -3^{\circ}\}$. 
That is, $M_{\text{model}}=3$.
The transition matrices of the three models $\bm{\mathrm{F}}_t$ are given by~\cite{dai2022adaptive}.
The initial model probability $\hat{\bm{\mathrm{u}}}_0^k=\bm{\mathrm{u}}_0^k=[0.8,0.1,0.1]'$, $k=1,\ldots,K$. 
The model transition probability matrix $\bm{\pi}^k$ of target $k$ is chosen as $[\bm{\pi}^k]_{i,j}=0.8$, if $i=j$ and $[\bm{\pi}^k]_{i,j}=0.1$, if $i\neq j$.
The rest simulation parameters are shown in Table~\ref{tab1}.
\begin{table}[!t]
\begin{center}
\caption{Simulation Parameters}
\label{tab1}
\begin{tabular}{ c | c | c | c | c | c }
\hline
Symbol & Value & Symbol & Value & Symbol & Value\\
\hline
$M_s$ & 2 & $N_s$ & 3 & $\sigma_{\phi}^2(0)$ & 0.002\\
\hline
$T$ & 1000 & $Q_s$ & 0.1 & $N_{\mathrm{pop}}$ & 50\\
\hline
$\bar{\alpha}$ & 0.2 & $\beta$ & 2 & $q_{\max}$ & 100\\
\hline
$\alpha$ & 0.998 & $\sigma_{r}^2(0)$ & 5 & $c_1=c_2$ & 2\\
\hline 
\end{tabular}
\end{center}
\end{table}

We consider the Best policy as a benchmark, where the optimal channel selections are obtained based on the ground-truth SINR, and compare MG-CRB-CL to it. 
In the $\epsilon$-greedy algorithm,
let $\epsilon = 0.1$, 
meaning that it explores random subsets in $10\%$ of the time. 
For the UCB1 algorithm,
let $\beta=2$.
For both the $\epsilon$-greedy and UCB1 algorithm,
the SINR state updates for target $k$ as
\begin{equation}\label{eq29}
\hat{Y}_{mn}^k(t)=\begin{cases}
\left(1-\bar{\alpha}^*\right)\hat{Y}_{mn}^k\left(t-1\right)+ \bar{\alpha}^*\gamma_{mn}^k\left(t\right), & \text{if played}\\
\hat{Y}_{mn}^k\left(t-1\right), & \text{otherwise.}
\end{cases} 
\end{equation}
where $\bar{\alpha}^*=0.998$.

We use the averaged RMSE (ARMSE) as the performance metric of target tracking accuracy, which is defined as ARMSE $\triangleq {1}/{T}$ $\sum_{t=1}^T \sqrt{1/{M_c}\sum_{m_c=1}^{M_c}\left[\left(\hat{{x}}_{t}\left(m_c\right)-x_t\right)^2+\left(\hat{{y}}_{t}\left(m_c\right)-y_t\right)^2 \right]}$,
where $\hat{{{x}}}_{t}(m_c)$ and $\hat{{{y}}}_{t}(m_c)$ are the estimated target location in $m_c$-th Monte-Carlo simulation, respectively, and $M_c$ is the number of Monte-Carlo trials.

To compare radar selection results between algorithms, define the average rate of the same selection with the Best policy, which is given for algorithm Alg
\begin{equation}\label{eq33}
\text{ASR}_{\mathrm{Alg}} \triangleq \frac{1}{M_c} \sum_{mc=1}^{M_c}\frac{1}{T} \sum_{t=1}^T\mathbb{I}\left(\bm{\mathcal{A}}_{\mathrm{Alg},t}=\bm{\mathcal{A}}_{ {\mathrm{Best}},t}\right),
\end{equation}
where $\mathbb{I}(\cdot)$ denotes the indicator function, which aims to find the same selection of algorithms for each instance with the Best policy. 
That is, $\text{ASR}_{\mathrm{Alg}}$ is calculated as a mismatch even though one radar station of the $\bm{\mathcal{A}}_{\mathrm{Alg},t}$ is different from the Best policy's selection $\bm{\mathcal{A}}_{\mathrm{Best},t}$. 

\subsection{Scenario 1: single target with NCV dynamic model}\label{subsec:IV-A}
In this section,
we evaluate the 
performance of MG-CRB-CL by tracking a single target. That is, $K=1$.
In this scenario, the target moves according to the NCV model over the time horizon $T$. 
The initial state of the target $\hat{\bm{\mathrm{X}}}_1=[5,100,5,100]'$ and 
the initial covariance matrix of the target state estimation error
$\hat{\bm{\mathrm{P}}}_1=\text{blkdiag}\left(20,20,20,20\right)$. 

\begin{figure*}[!t]
    \begin{minipage}[t]{0.33\linewidth}
        \centering
        \includegraphics[width=2.0in]{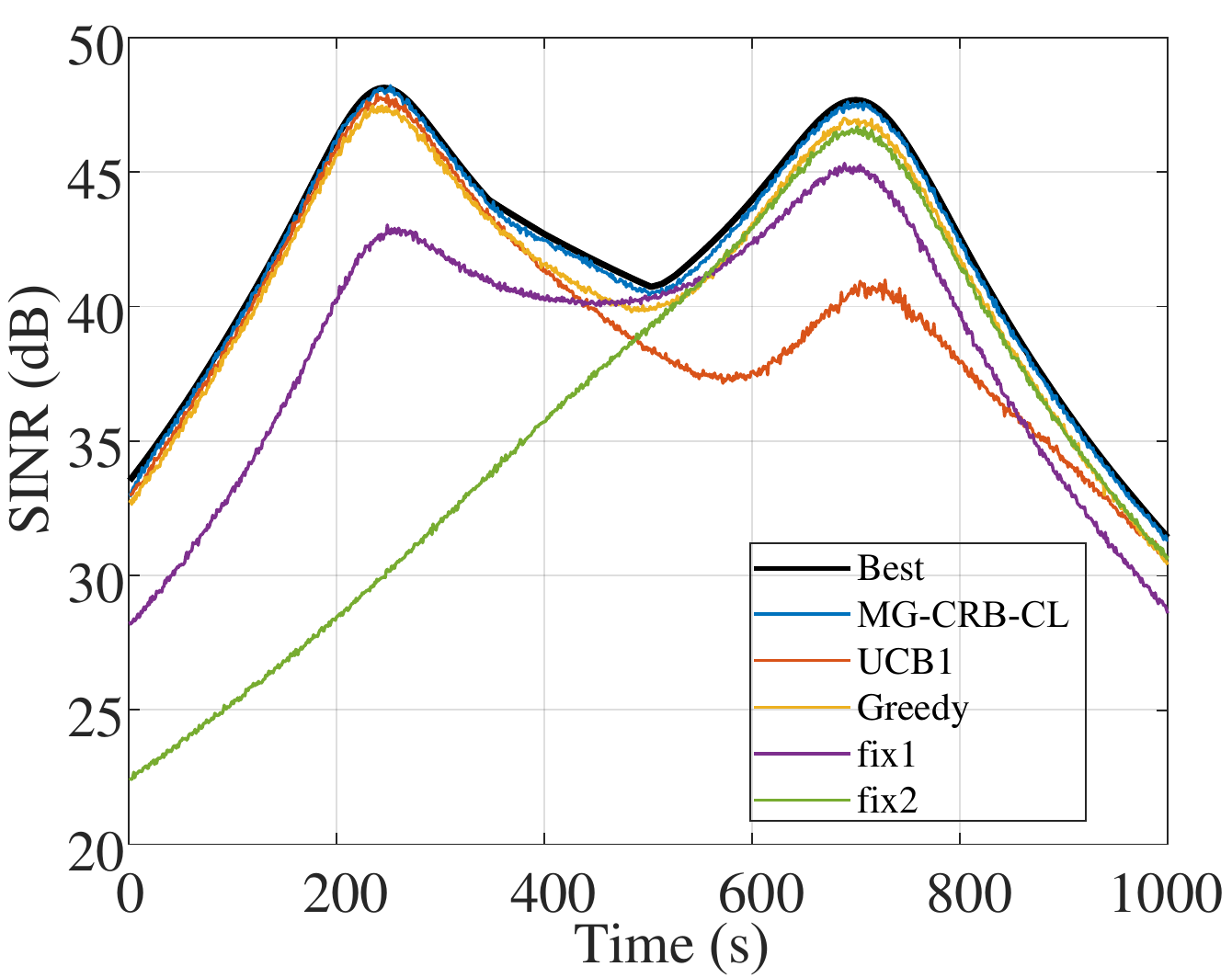}
        \centerline{(a)}
    \end{minipage}%
    \begin{minipage}[t]{0.33\linewidth}
        \centering
        \includegraphics[width=2.05in]{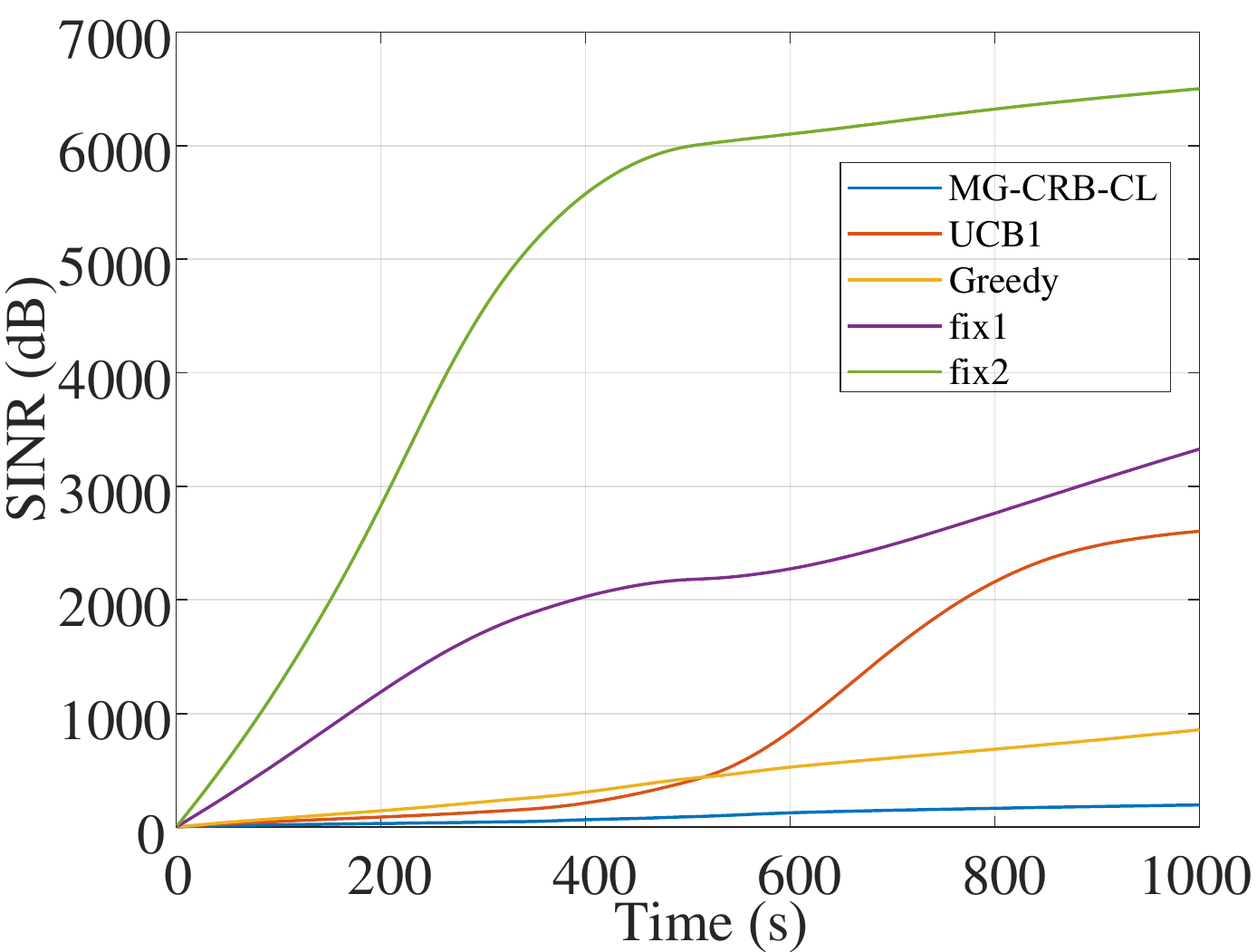}
        \centerline{(b)}
    \end{minipage}
    \begin{minipage}[t]{0.33\linewidth}
        \centering
        \includegraphics[width=2.0in]{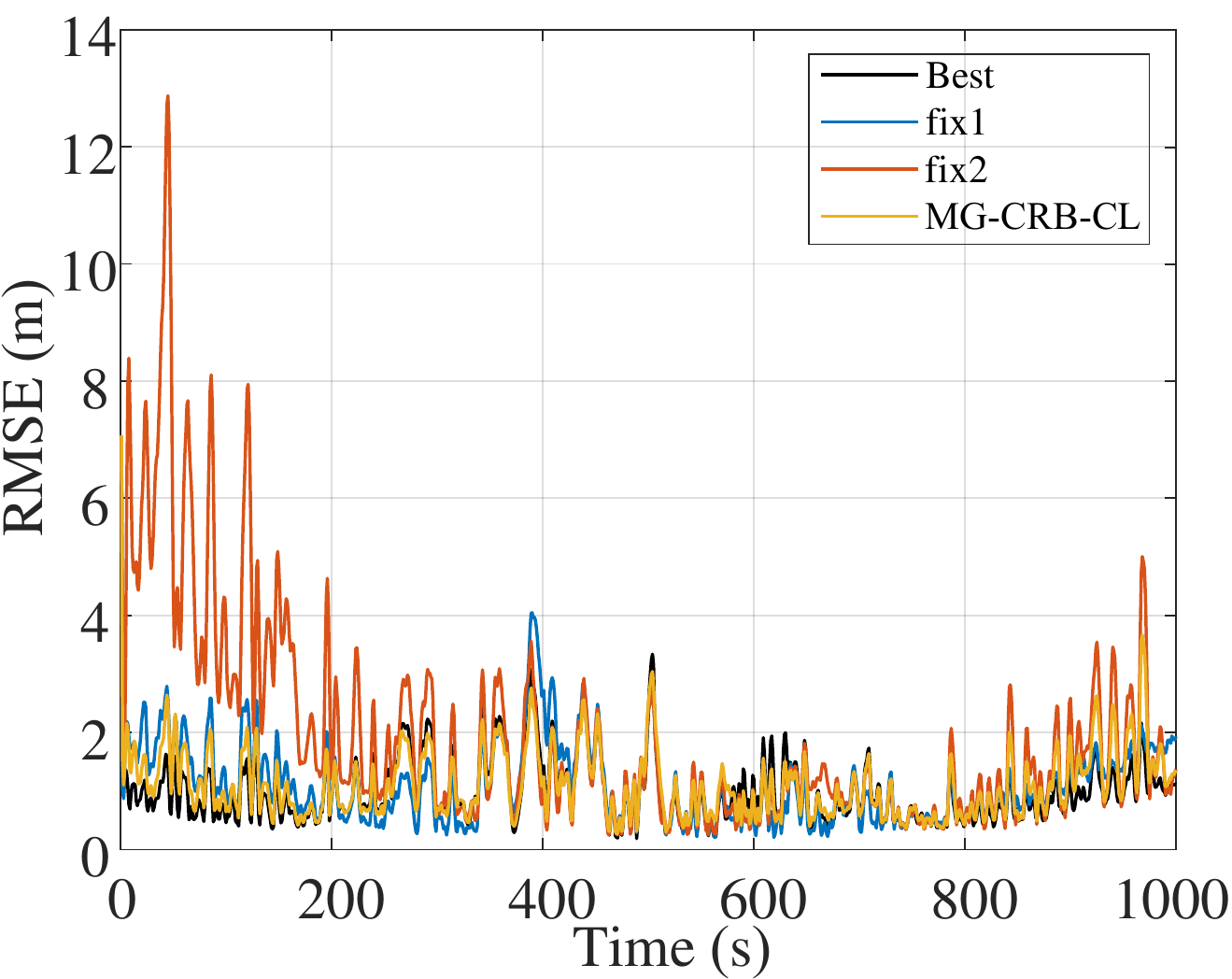}
        \centerline{(c)}
    \end{minipage}
    \caption{Performance metrics for each algorithm in Scenario 1. (a) Average SINR (dB). (b) Regret SINR (dB). (c) RMSE (m).}
    \label{fig_sim7}
\end{figure*}

Fig.~\ref{fig_sim7}(a) shows the average SINR. 
By the relative geometry in Fig.~\ref{fig_sim5}, the ``fix1" policy is a relatively balanced solution over the entire horizon, while the ``fix2" policy may be only superior in the later period. 
The two policies obtain higher SINR values in the first half and second half of the time horizon, respectively. 
Compared with fixed policies, the $\epsilon$-greedy algorithm obtains higher SINRs all the time, 
while the UCB1 algorithm only achieves a better performance in the time period 0$-$400 s.
This may be caused by linear error accumulation from \eqref{eq29} without updates of SINR measurements and indicates that UCB1 is not appropriate for the case with moving targets.
The MG-CRB-CL algorithm achieves the highest performance,
with consistently the highest SINR values close to the maximum all the time, and has almost the same performance as the Best policy. 
The regret in Fig.~\ref{fig_sim7}(b) also shows that MG-CRB-CL obtains the lowest SINR regret. 
The UCB1 algorithm obtains the lower regret than the $\epsilon$-greedy algorithm before time $t=455$~s, and vice versa after time $t=455$~s.
The two fixed policies achieve the highest regrets compared with the other three algorithms. 

In Fig.~\ref{fig_sim7}(c), we consider the RMSE metric as a validation of the tracking performance.
The ``fix1" policy derives the lowest RMSE value only in time 268$-$360 s; 
the ``fix2" policy indeed obtains the worst tracking performance in most of the time, except for the period 500$-$590 s. 
All in all, MG-CRB-CL achieves the best RMSE.
The ARMSE of all the six policies over the time horizon $T$ is summarized in Table~\ref{tab2}, which further proves the effectiveness of MG-CRB-CL. 
\begin{table}[!t]
\begin{center}
\caption{Comparison of ARMSE (m) through the Time Horizon \textit{T}}
\label{tab2}
\begin{tabular}{ c | c | c | c }
\hline
{Algorithms} & {ARMSE} & {Algorithms} & {ARMSE} \\
\hline
Best & 0.9863 & MG-CRB-CL & 1.1030 \\
\hline
UCB1 & 1.3633 & $\epsilon$-greedy & 1.2353 \\
\hline
fix1 & 1.1284 & fix2 & 1.9089 \\ 
\hline 
\end{tabular}
\end{center}
\end{table}

\begin{figure}[!t]
    \begin{minipage}[t]{0.5\linewidth}
        \centering
        \includegraphics[width=1.55in]{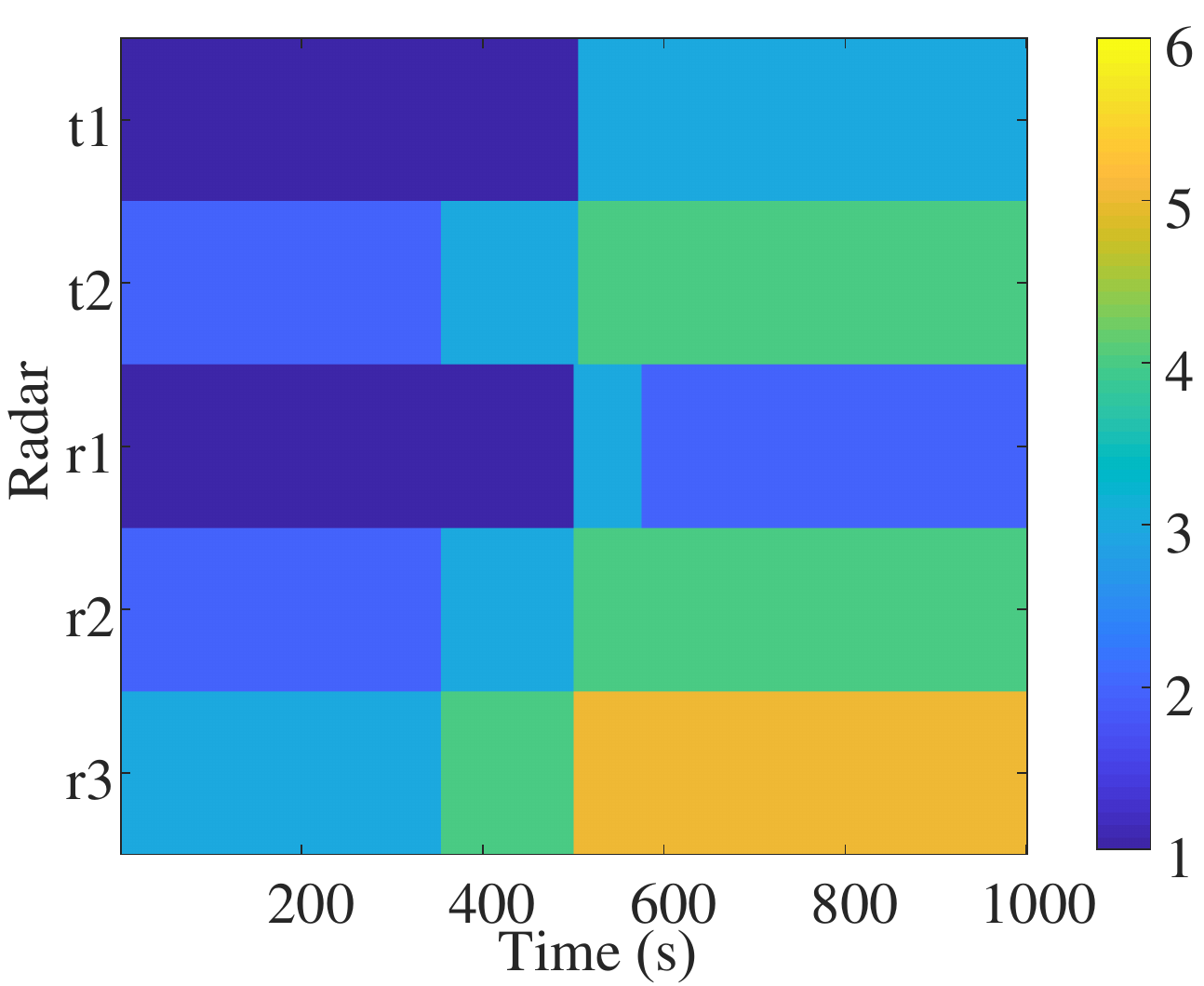}
        \centerline{(a)}
    \end{minipage}%
    \begin{minipage}[t]{0.5\linewidth}
        \centering
        \includegraphics[width=1.55in]{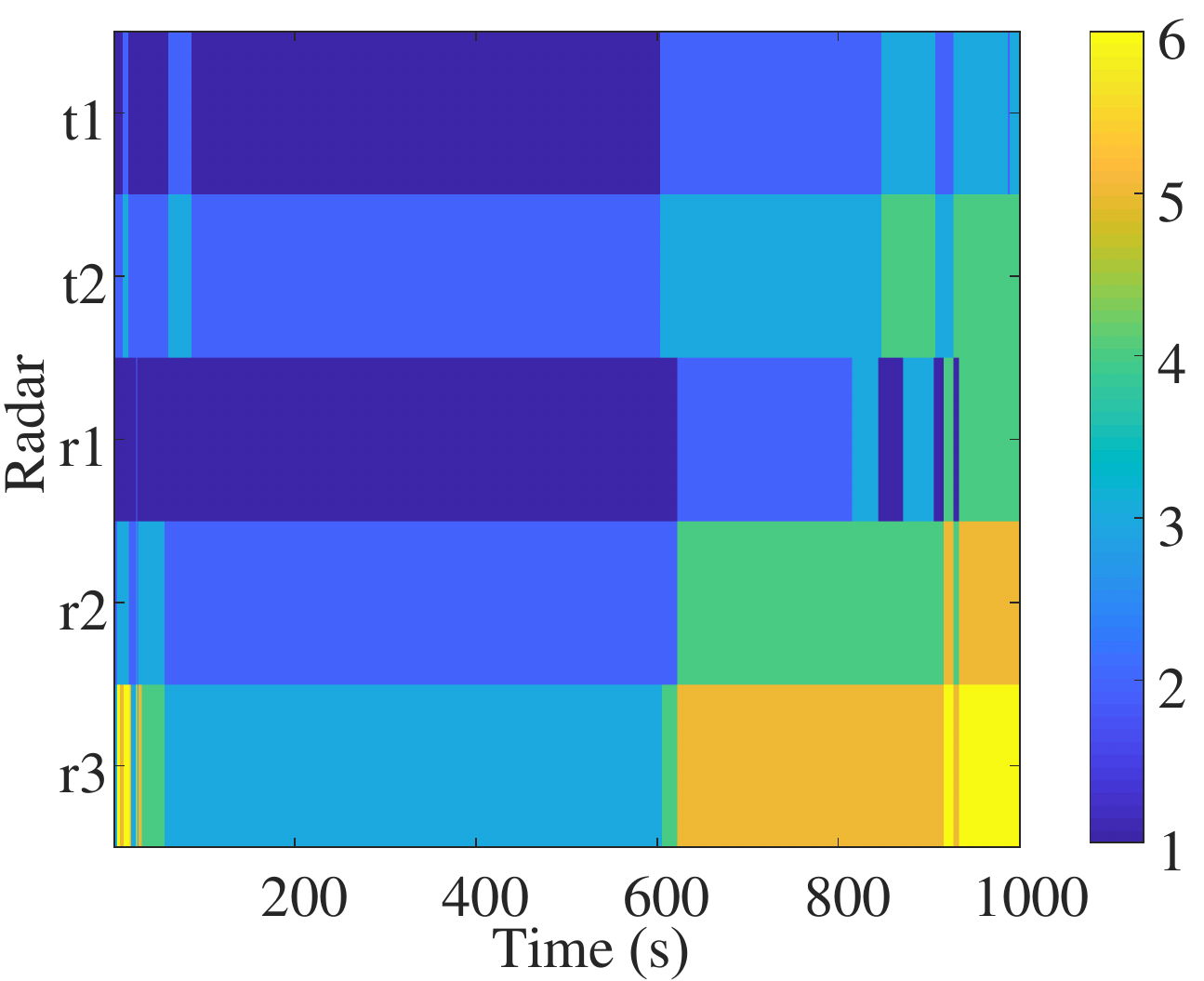}
        \centerline{(b)}
    \end{minipage}
    \begin{minipage}[t]{0.5\linewidth}
        \centering
        \includegraphics[width=1.55in]{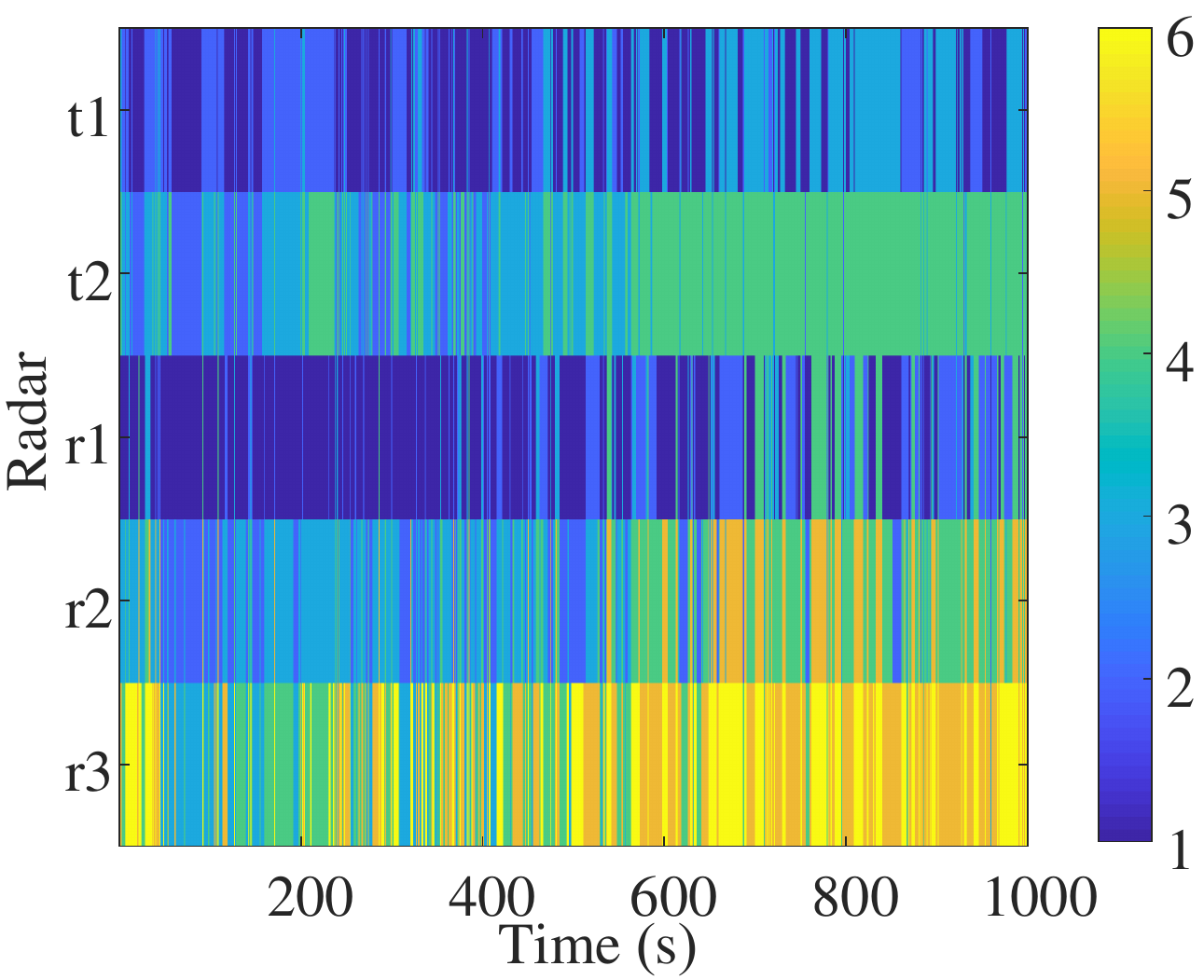}
        \centerline{(c)}
    \end{minipage}%
    \begin{minipage}[t]{0.5\linewidth}
        \centering
        \includegraphics[width=1.55in]{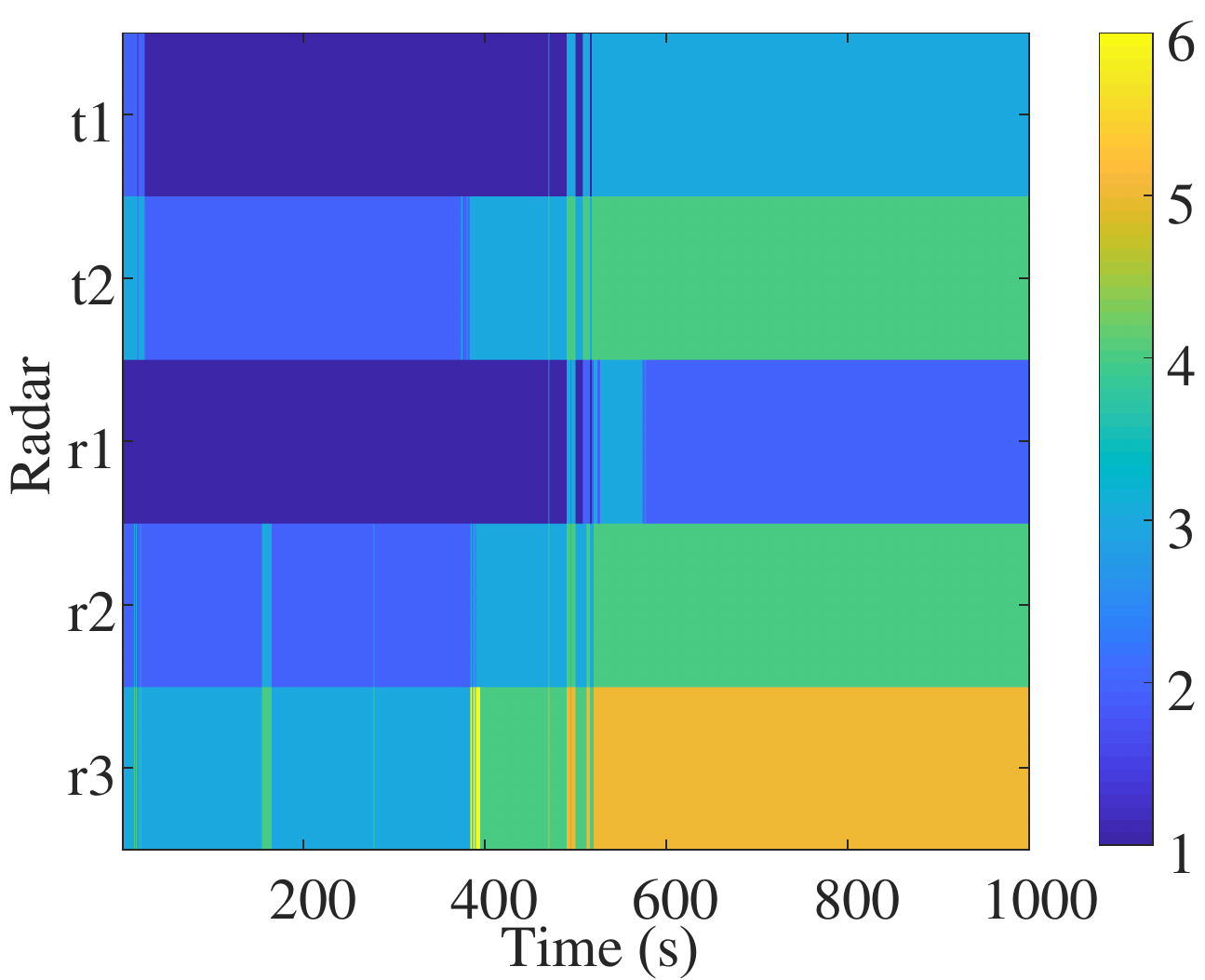}
        \centerline{(d)}
    \end{minipage}%
    \caption{Radar station selection index. (a) Best policy. (b) UCB1 algorithm. (c) $\epsilon$-greedy algorithm. (d) MG-CRB-CL algorithm.}
    \label{fig_sim9}
\end{figure}

Finally, Fig.~\ref{fig_sim9} demonstrates the selection results $\bm{\mathcal{A}}_{t}$ via color maps, where $\lbrace\text{`t1', `t2', `r1', `r2', `r3'}\rbrace$ in the $y$-axis denote the two selected transmitters and three selected receivers, respectively. The color value represents the \textit{m}-th transmitter or the \textit{n}-th receiver. 
In Fig.~\ref{fig_sim9}(a), the Best policy selects the subset $\lbrace\left[1,2\right],\left[1,2,3\right]\rbrace$ in time 0$-$380 s and the subset $\lbrace\left[3,4\right],\left[2,4,5\right]\rbrace$ in time 590$-$1000 s, which is consistent with the geometry between the target and the MIMO radar.
Compared with  $\bm{\mathcal{A}}_{\mathrm{Best},t}$ for the Best policy, UCB1 obtains the average rate of the same selection in $\text{ASR}=7.63\%$, while $\epsilon$-greedy obtains $\text{ASR}=14.56\%$. 
MG-CRB-CL obtains the best performance in $\text{ASR}=45.84\%$, delivering at least three times better performance than that of the other algorithms. The absolute value of ASR may not be close to 100\%, which is resulted from the strong indication condition in \eqref{eq33}.
Namely, when $\bm{\mathcal{A}}_{\text{MG-CRB-CL}}=\{[3,4],[3,4,5]\}$ and $\bm{\mathcal{A}}_{\mathrm{Best}}=\{[3,4],[2,4,5]\}$, then $\mathbb{I}\left(\bm{\mathcal{A}}_{\text{MG-CRB-CL}}=\bm{\mathcal{A}}_{\mathrm{Best}}\right)=0$.
MG-CRB-CL selects the same radar stations with the Best policy in most time slots and achieves near-optimal performance.  

\subsection{Scenario 2: single target with NCV and NCT dynamic models}\label{subsec:IV-B}
This section again evaluates the performance of MG-CRB-CL via tracking a single target. 
The target dynamic models used in this scenario are NCV and NCT over the time horizon $T$.
As shown in Fig.~\ref{fig_sim5}(b), the starting point of the target is the red dot. 
The initial target dynamic state $\hat{\bm{\mathrm{X}}}_1=\left[5005,100,25005,100\right]'$. 
The target moves following the NCT dynamic model in time 400$-$440 s with $\omega=3^{\circ}$ and in time 600$-$640 s with $\omega=-3^{\circ}$, respectively. 
The estimation error covariance matrix is initialized with $\hat{\bm{\mathrm{P}}}_1=\text{blkdiag}\left(20,20,20,20\right)$. 

The parameter settings of UCB1 and $\epsilon$-greedy remain the same as those in Scenario 1.
The SINR state update equation is established by using \eqref{eq29}.

\begin{figure*}[!t]
    \begin{minipage}[t]{0.33\linewidth}
        \centering
        \includegraphics[width=2.0in]{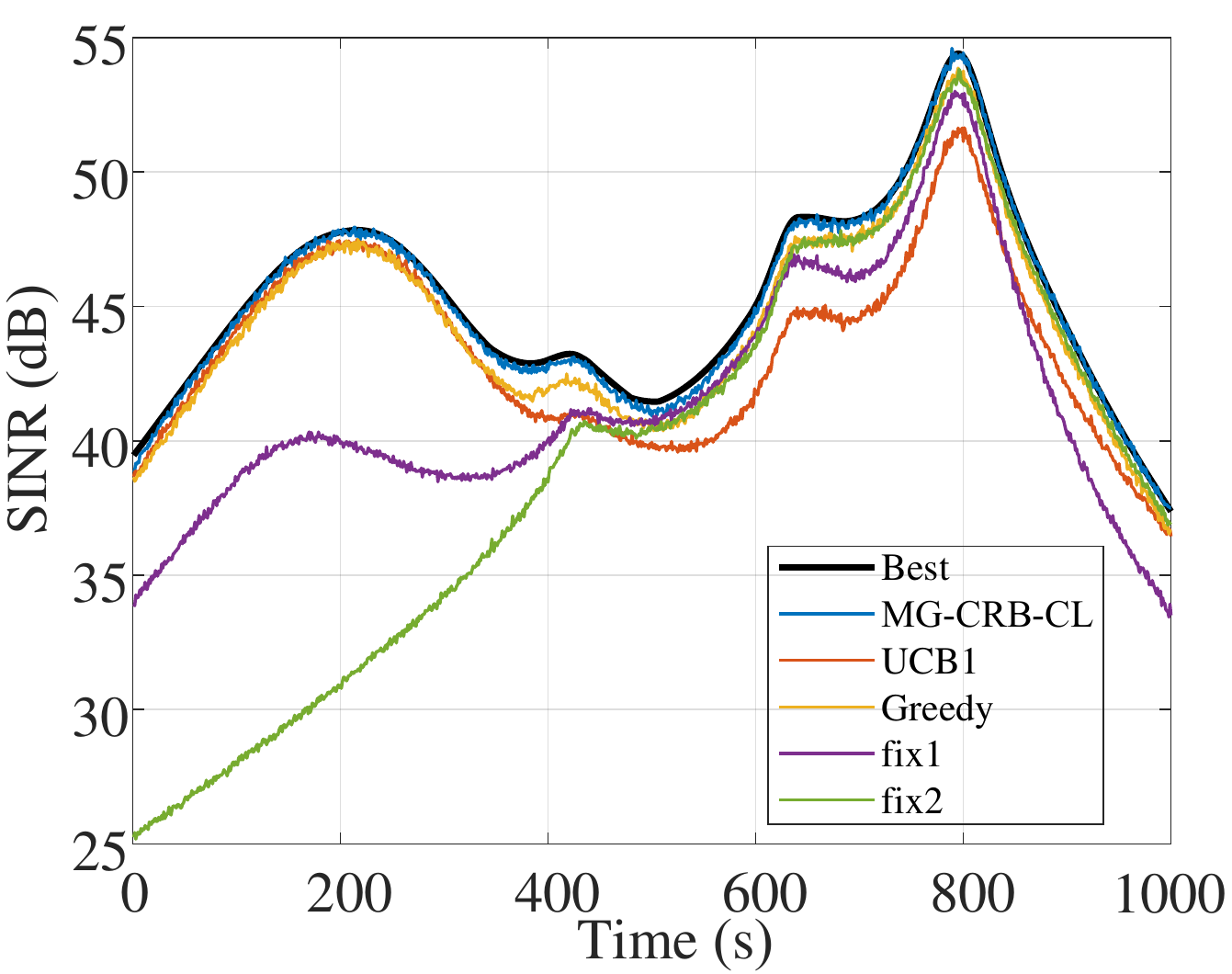}
        \centerline{(a)}
    \end{minipage}%
    \begin{minipage}[t]{0.33\linewidth}
        \centering
        \includegraphics[width=2.05in]{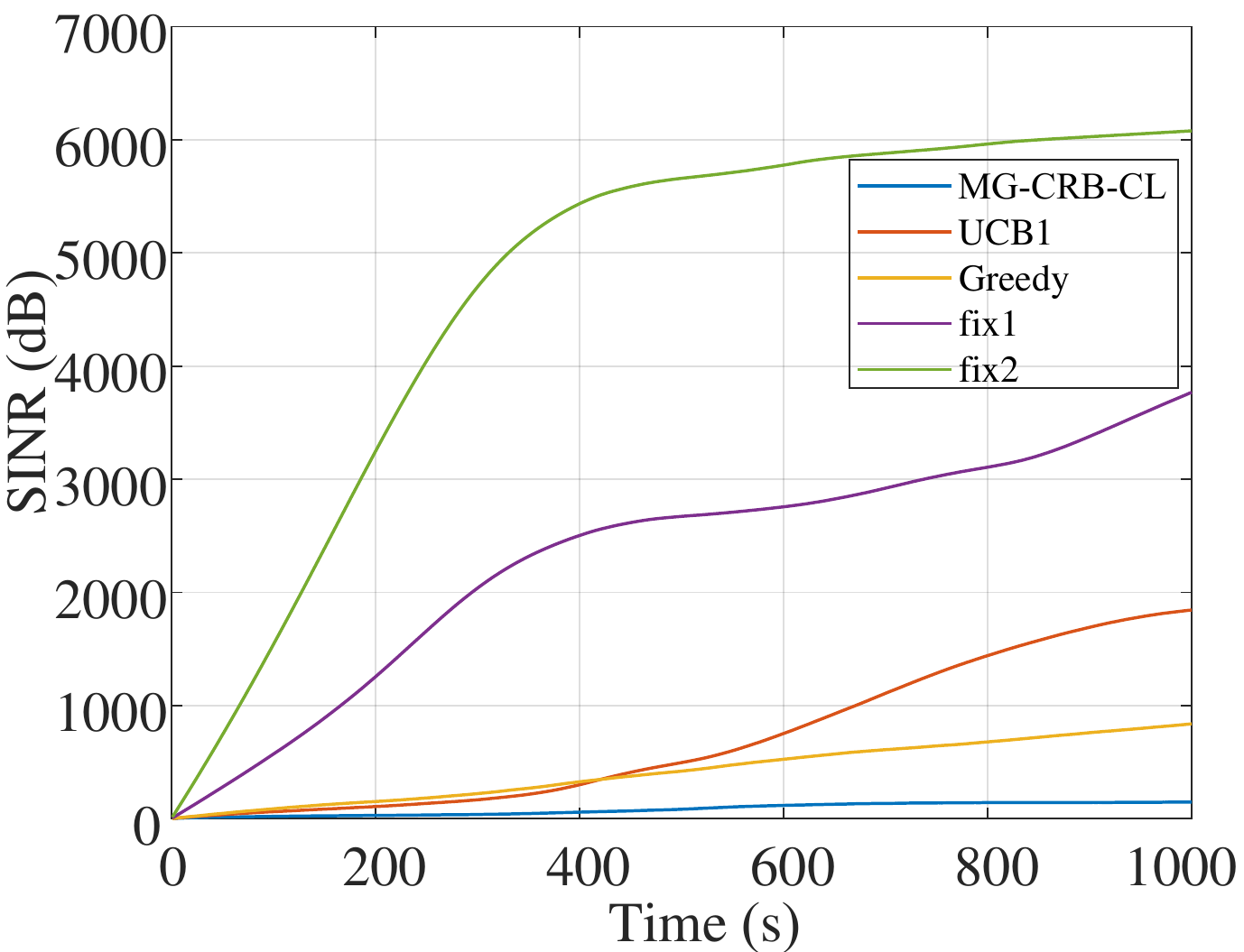}
        \centerline{(b)}
    \end{minipage}
    \begin{minipage}[t]{0.33\linewidth}
        \centering
        \includegraphics[width=2.0in]{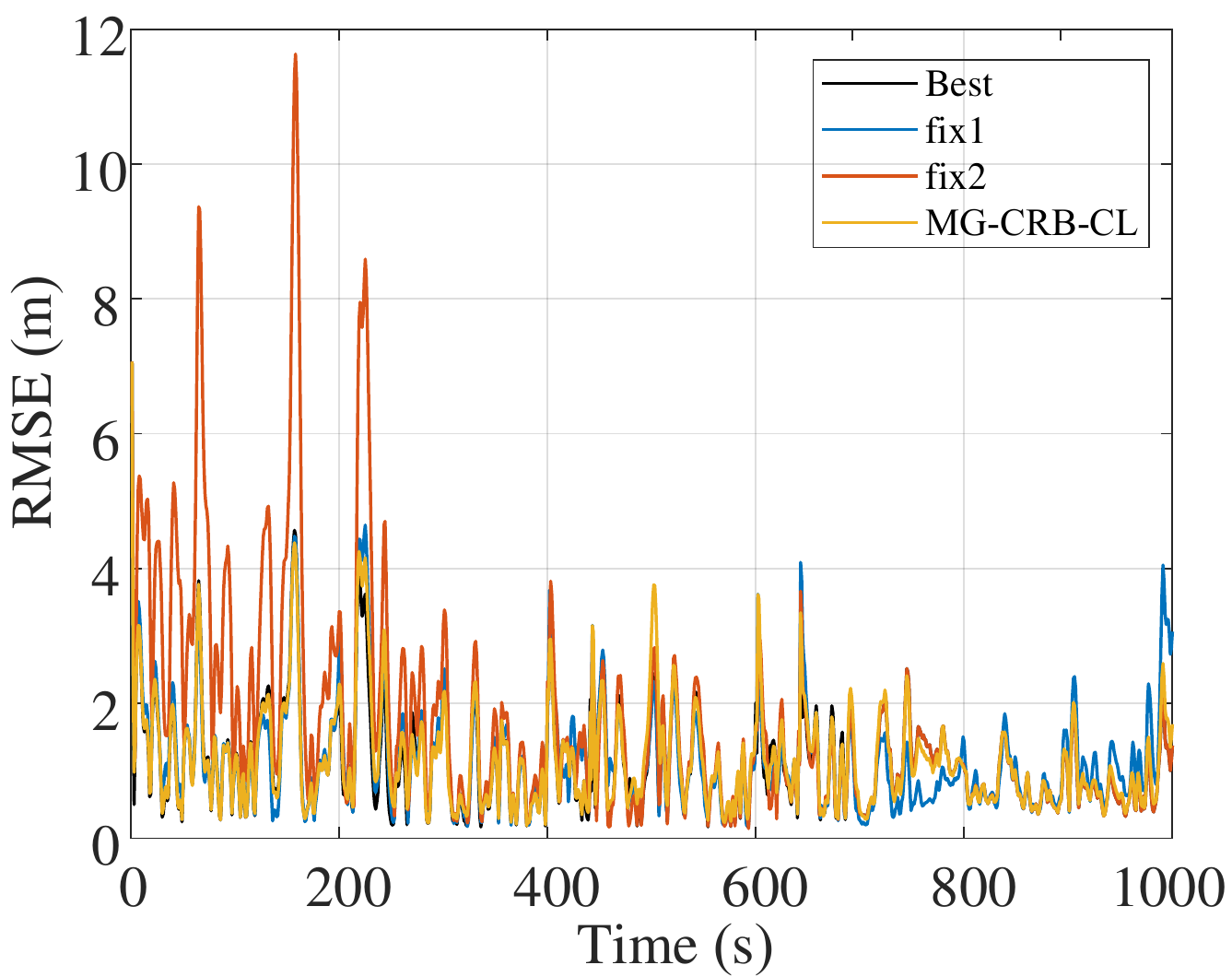}
        \centerline{(c)}
    \end{minipage}
    \caption{Performance metrics for each algorithm in Scenario 2. (a) Average SINR (dB). (b) Regret SINR (dB). (c) RMSE (m).}
    \label{fig_sim11}
\end{figure*}

The average SINR is shown in Fig.~\ref{fig_sim11}(a). 
Similar to Scenario 1, $\epsilon$-greedy is superior to the two fixed selection policies, particularly in the time period of 0 to 500 s, while the performance of UCB1 deteriorates after the time 420 s. 
After the target maneuvers in time $t=400$, apart from MG-CRB-CL,
all the optimization policies exhibit performance degradation.
In  Fig.~\ref{fig_sim11}(b), compared with MG-CRB-CL, the benchmark algorithms again have worse performances and higher regrets.

The tracking performance is validated by the RMSE metric, as shown in Fig.~\ref{fig_sim11}(c). MG-CRB-CL achieves the best RMSE performance in most of the time, indicating that the online selection scheme of MG-CRB-CL significantly improves the tracking performance, even for maneuvering targets. 
It highlights the significance of MG-CRB-CL in achieving better optimization objectives than the algorithms with fixed selections. 
The results of ARMSE over the time horizon $T$ is presented in Table~\ref{tab3}, where MG-CRB-CL achieves the lowest RMSE. 
It further enhances the effectiveness and advantages of MG-CRB-CL, even in the case of maneuvering targets. 
\begin{table}[!t]
\begin{center}
\caption{Comparison of ARMSE (m) through the Time Horizon \textit{T}}
\label{tab3}
\begin{tabular}{ c | c | c | c }
\hline
{Algorithms} & {ARMSE} & {Algorithms} & {ARMSE} \\
\hline
Best & 1.1242 & MG-CRB-CL & 1.1721 \\
\hline
UCB1 & 1.3159 & $\epsilon$-greedy & 1.2230 \\
\hline
fix1 & 1.1835 & fix2 & 1.7010 \\ 
\hline 
\end{tabular}
\end{center}
\end{table}

\begin{figure}[!t]
    \begin{minipage}[t]{0.5\linewidth}
        \centering
        \includegraphics[width=1.55in]{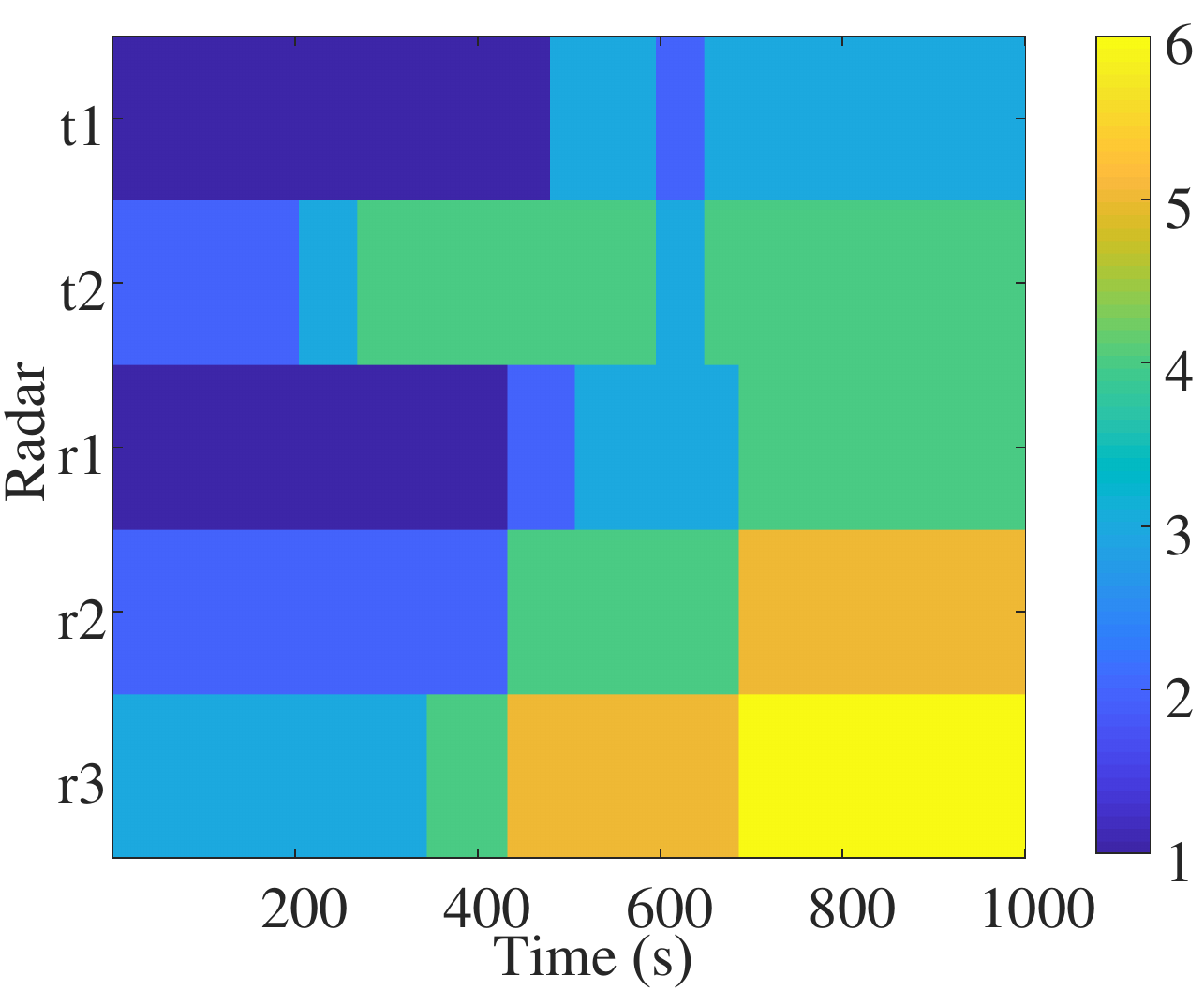}
        \centerline{(a)}
    \end{minipage}%
    \begin{minipage}[t]{0.5\linewidth}
        \centering
        \includegraphics[width=1.55in]{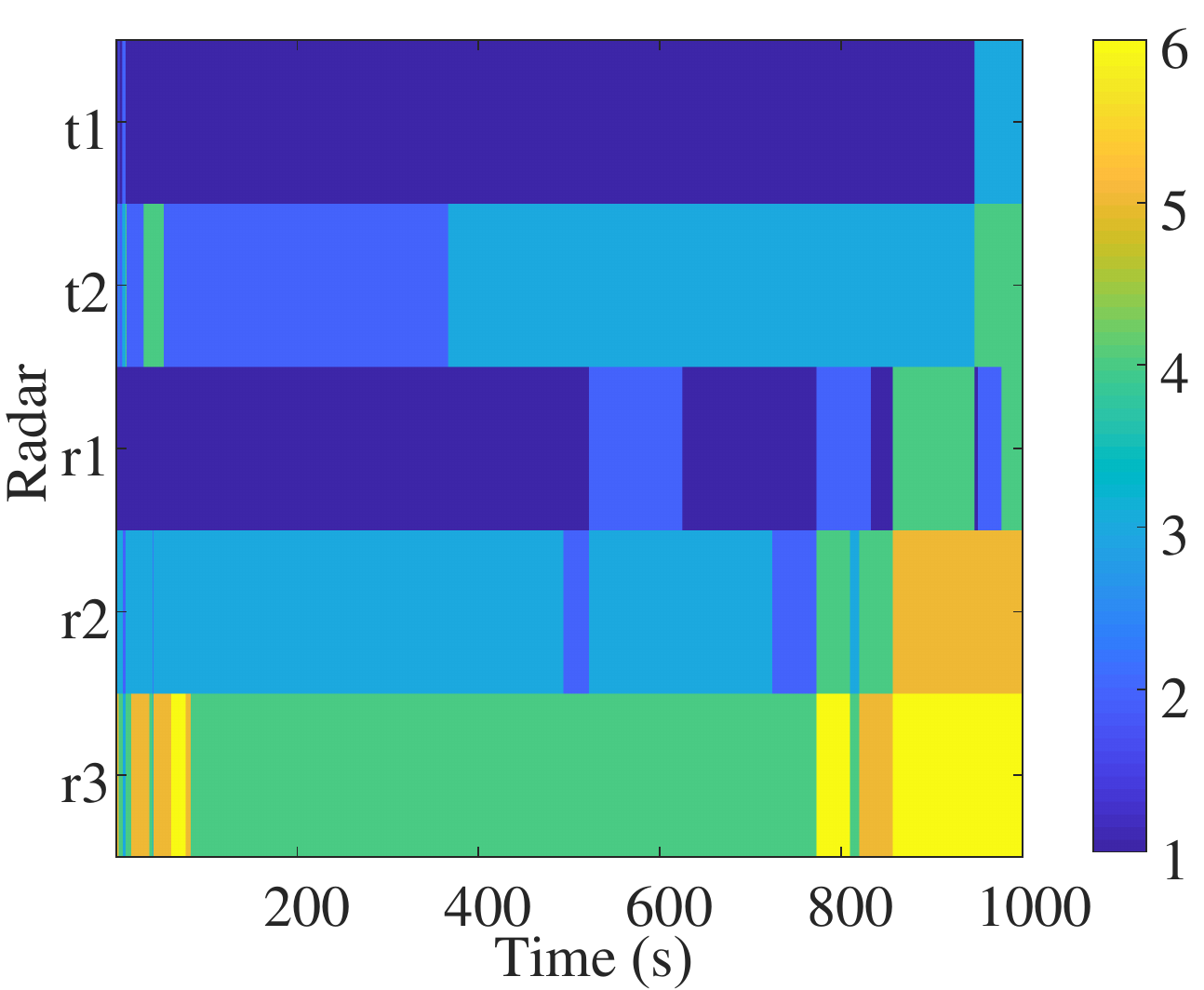}
        \centerline{(b)}
    \end{minipage}
    \begin{minipage}[t]{0.5\linewidth}
        \centering
        \includegraphics[width=1.55in]{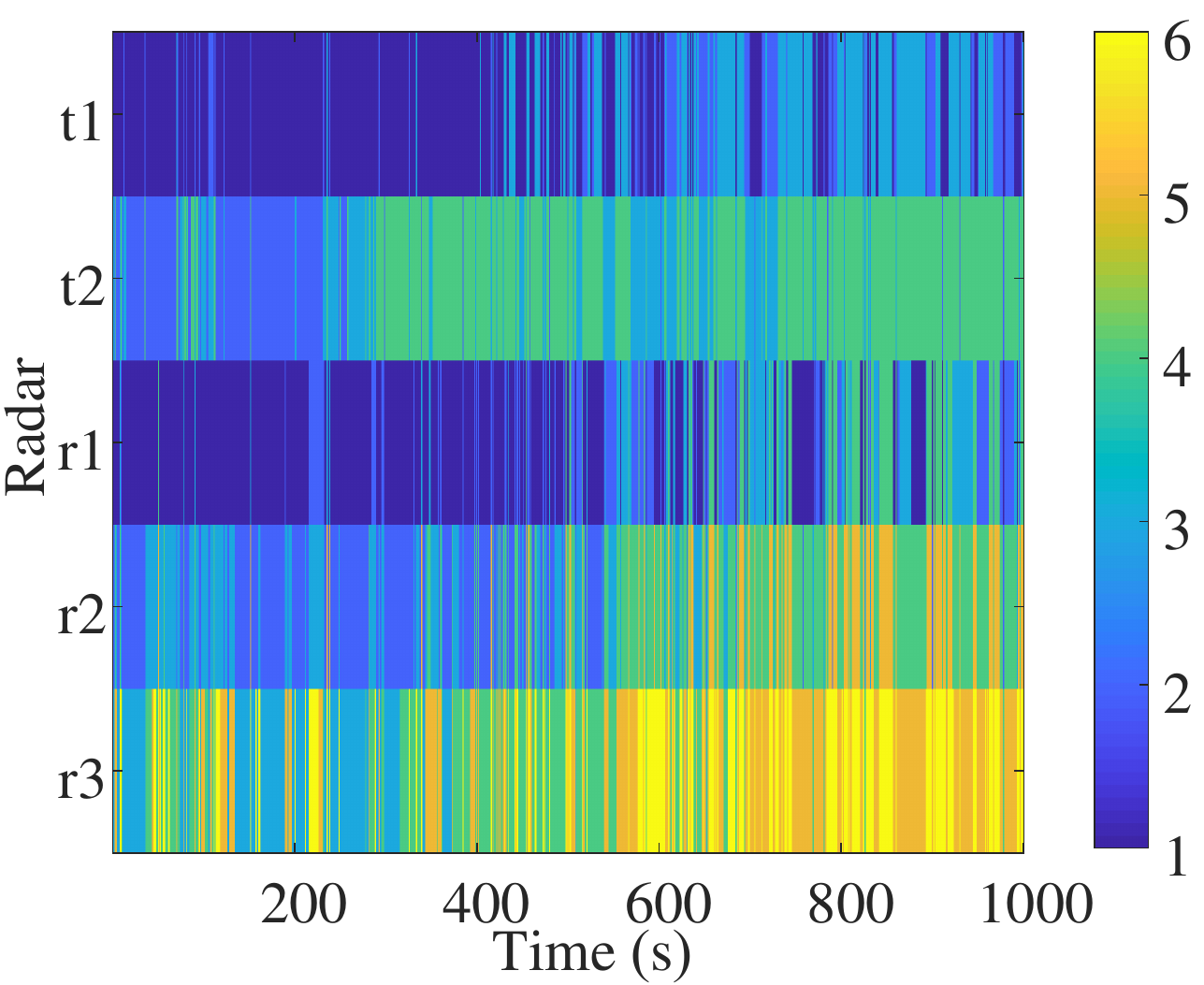}
        \centerline{(c)}
    \end{minipage}%
    \begin{minipage}[t]{0.5\linewidth}
        \centering
        \includegraphics[width=1.55in]{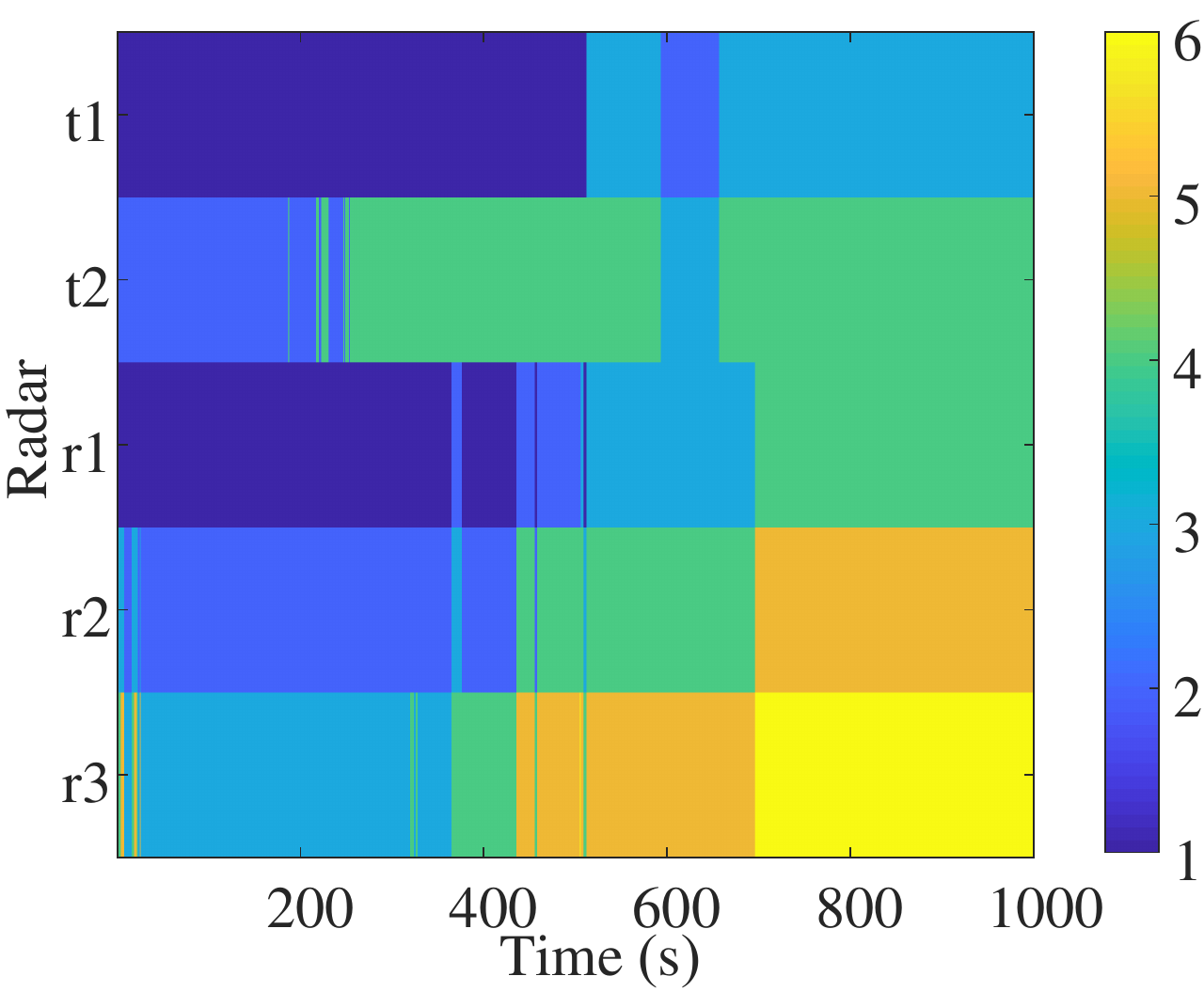}
        \centerline{(d)}
    \end{minipage}%
    \caption{Radar station selection index. (a) Best policy. (b) UCB1 algorithm. (c) $\epsilon$-greedy algorithm. (d) MG-CRB-CL algorithm.}
    \label{fig_sim14}
\end{figure}

The selection results are shown  in Fig.~\ref{fig_sim14} with color maps. The color values indicate that MG-CRB-CL selects the same radar stations as the Best policy in most of the time with an ASR $=56.66\%$; while the UCB1 and $\epsilon$-greedy algorithms act sub-optimal selections with ASR metric being $11.45\%$ and $20.51\%$, respectively.

\subsection{Scenario 3: multiple targets with multiple dynamic models}\label{subsec:IV-C}
This section evaluates the performance of MG-CRB-CL via tracking multiple targets. 
Based on Scenario 2, in this scenario, we add a second target moving according to an NCV model over the time horizon $T$.
Target 1 moves according to the NCT dynamic model in time slot 400 s and 600 s, and according to the NCV model for all the remaining time periods. 
As shown in Fig.~\ref{fig_sim5}(c), the starting points of the targets are in the red dots. 
The initial target dynamic states of targets $\hat{\bm{\mathrm{X}}}_{1}^1=\left[5010,100,25010,100\right]'$ and $\hat{\bm{\mathrm{X}}}_{1}^2=\left[10,50,100010,0\right]'$.
The estimation error covariance matrices are both initialized with $\hat{\bm{\mathrm{P}}}_{1}^1=\hat{\bm{\mathrm{P}}}_{1}^2=\text{blkdiag}\left(20,20,20,20\right)$. 

Considering the difficulty of tracking maneuvering Target 1, define the weights of targets to reflect the relative importance of tracking each target in the optimization objective of MG-CRB-CL. Consider 
$\Omega_1=0.6$ and $\Omega_2=0.4$.

\begin{figure*}[!t]
\centering
    \begin{minipage}[t]{0.33\linewidth}
        \centering
        \includegraphics[width=2.0in]{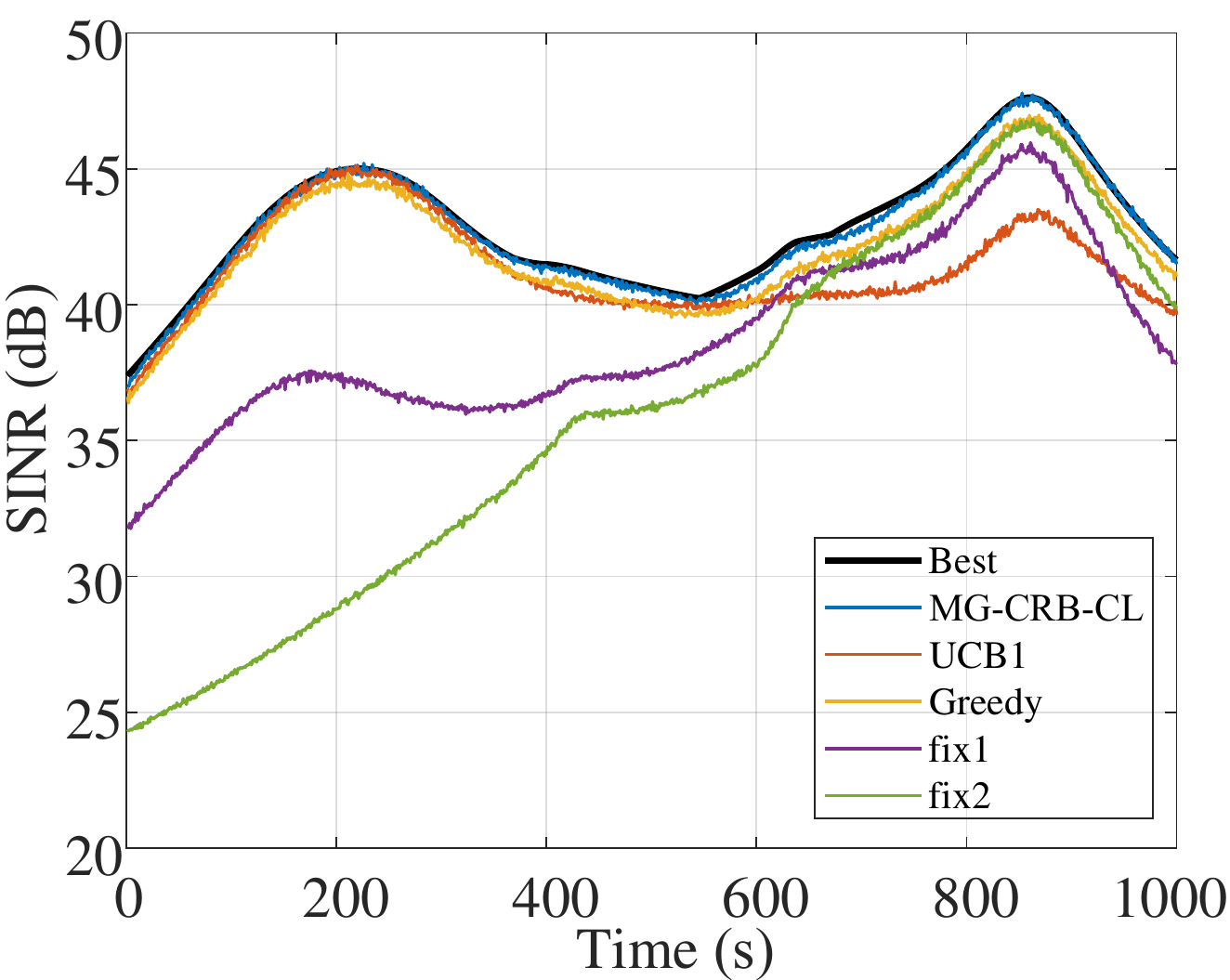}
        \centerline{(a)}
    \end{minipage}%
    \begin{minipage}[t]{0.33\linewidth}
        \centering
        \includegraphics[width=2.1in]{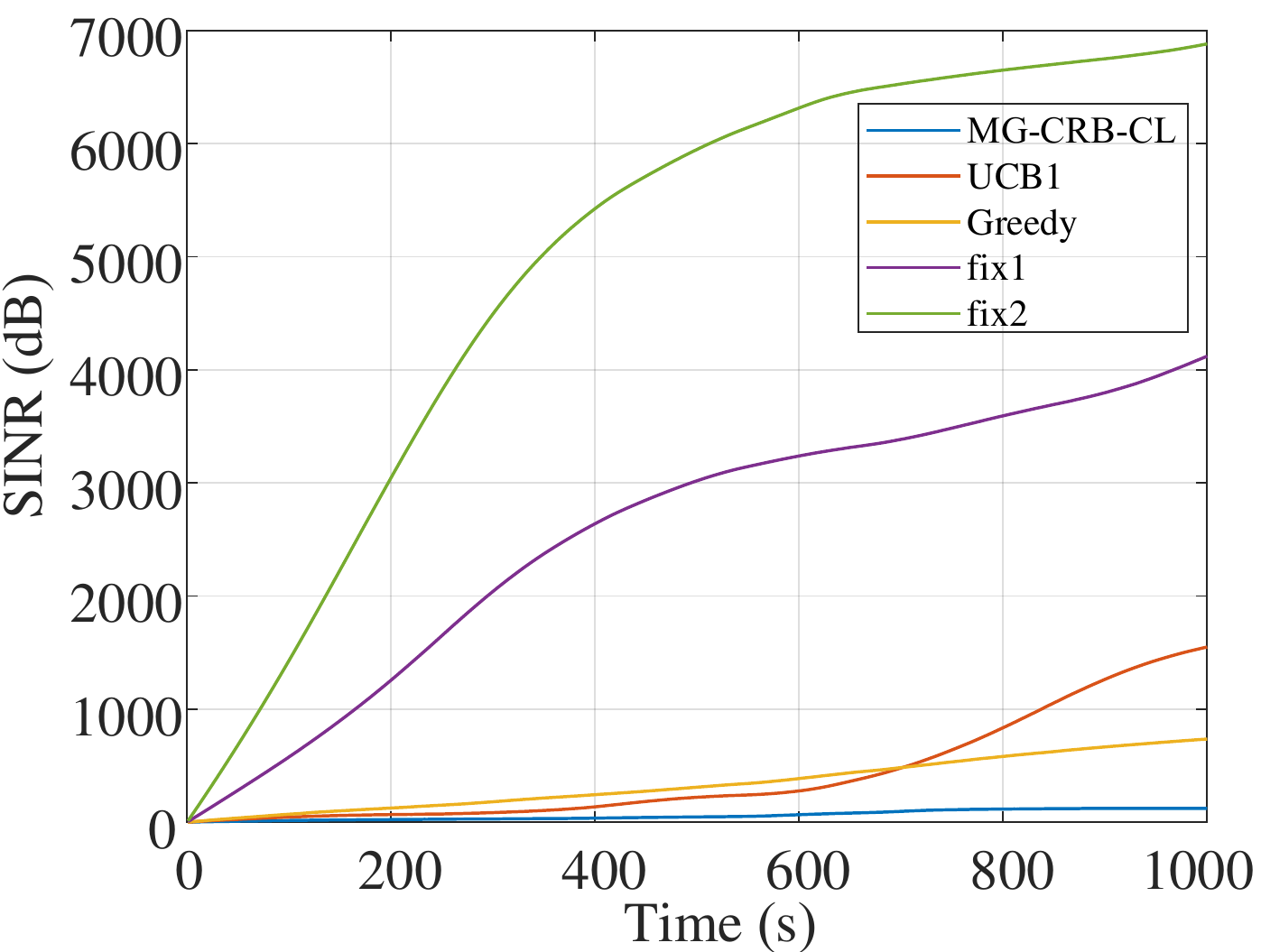}
        \centerline{(b)}
    \end{minipage}
    \caption{Performance metrics for each algorithm in Scenario 3. (a) Average SINR (dB). (b) Regret SINR (dB).}
    \label{fig_sim16}
\end{figure*}

The average SINR performance is presented in Fig.~\ref{fig_sim16}(a). 
In Fig.~\ref{fig_sim16}(a), observe that MG-CRB-CL outperforms all the other algorithms with respect to the weighted fusion objective function and achieves the lowest regret.

\begin{figure*}[!t]
\centering
    \begin{minipage}[t]{0.33\linewidth}
        \centering
        \includegraphics[width=2.0in]{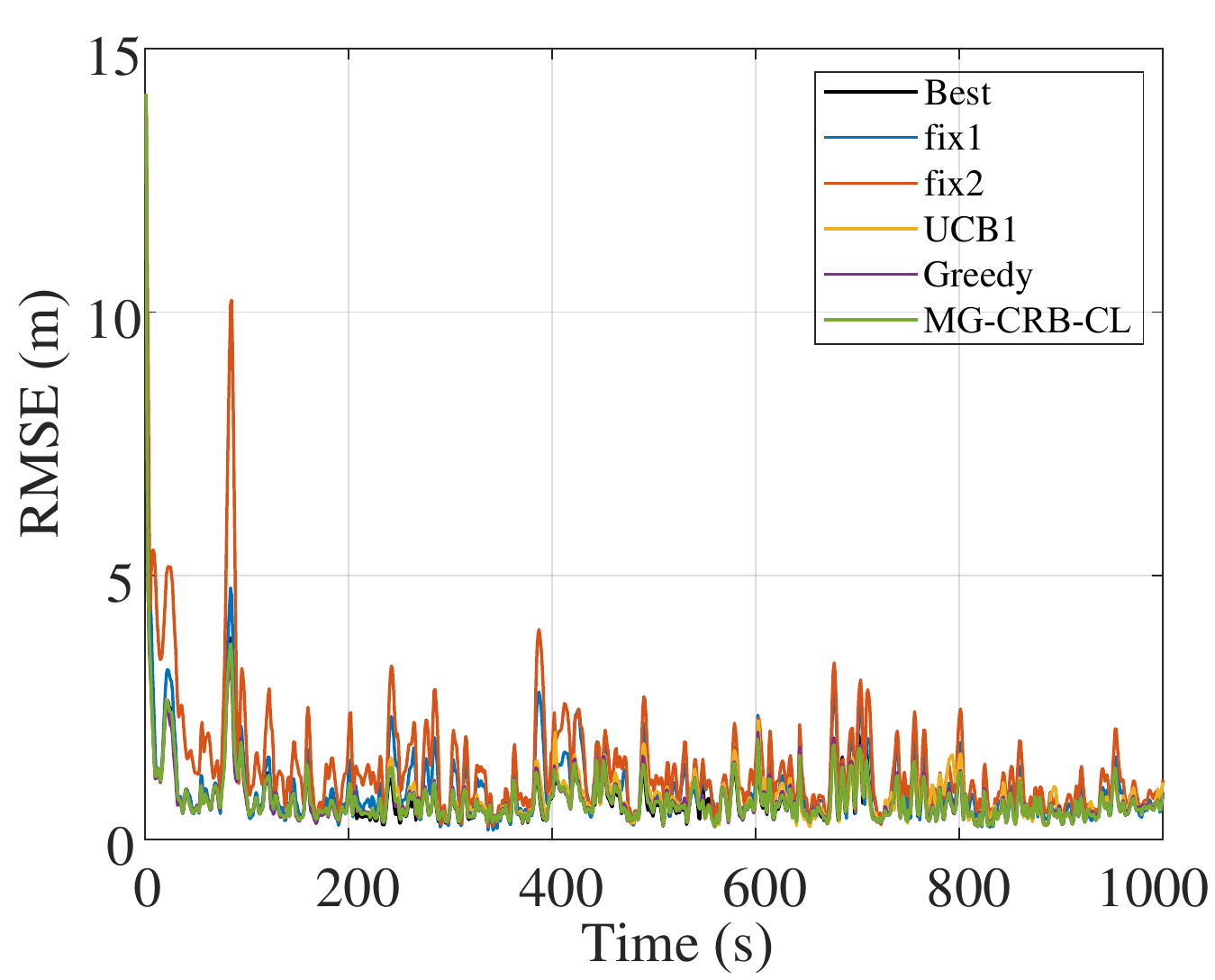}
        \centerline{(a)}
    \end{minipage}%
    \begin{minipage}[t]{0.33\linewidth}
        \centering
        \includegraphics[width=2.0in]{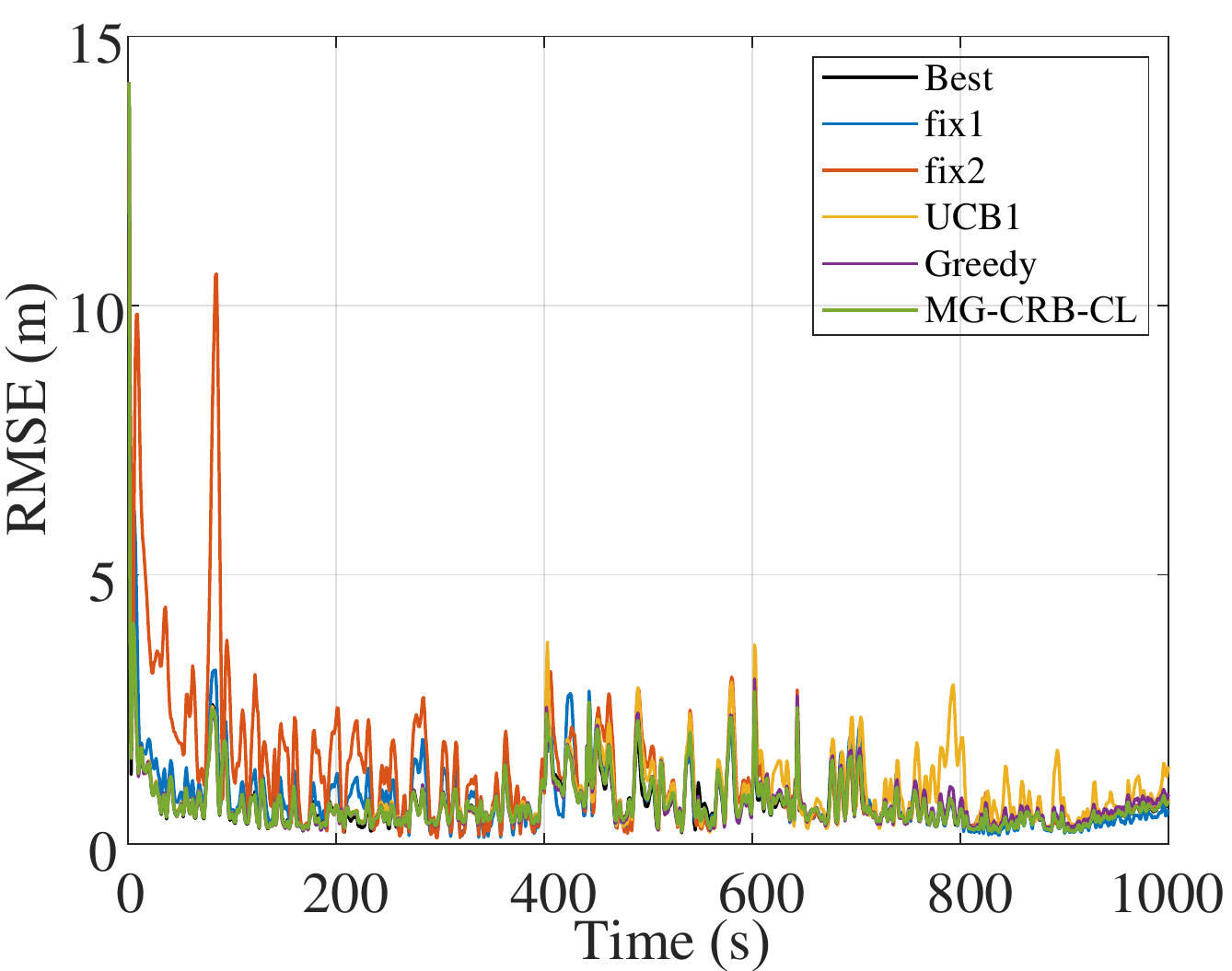}
        \centerline{(b)}
    \end{minipage}
    \begin{minipage}[t]{0.33\linewidth}
        \centering
        \includegraphics[width=2.0in]{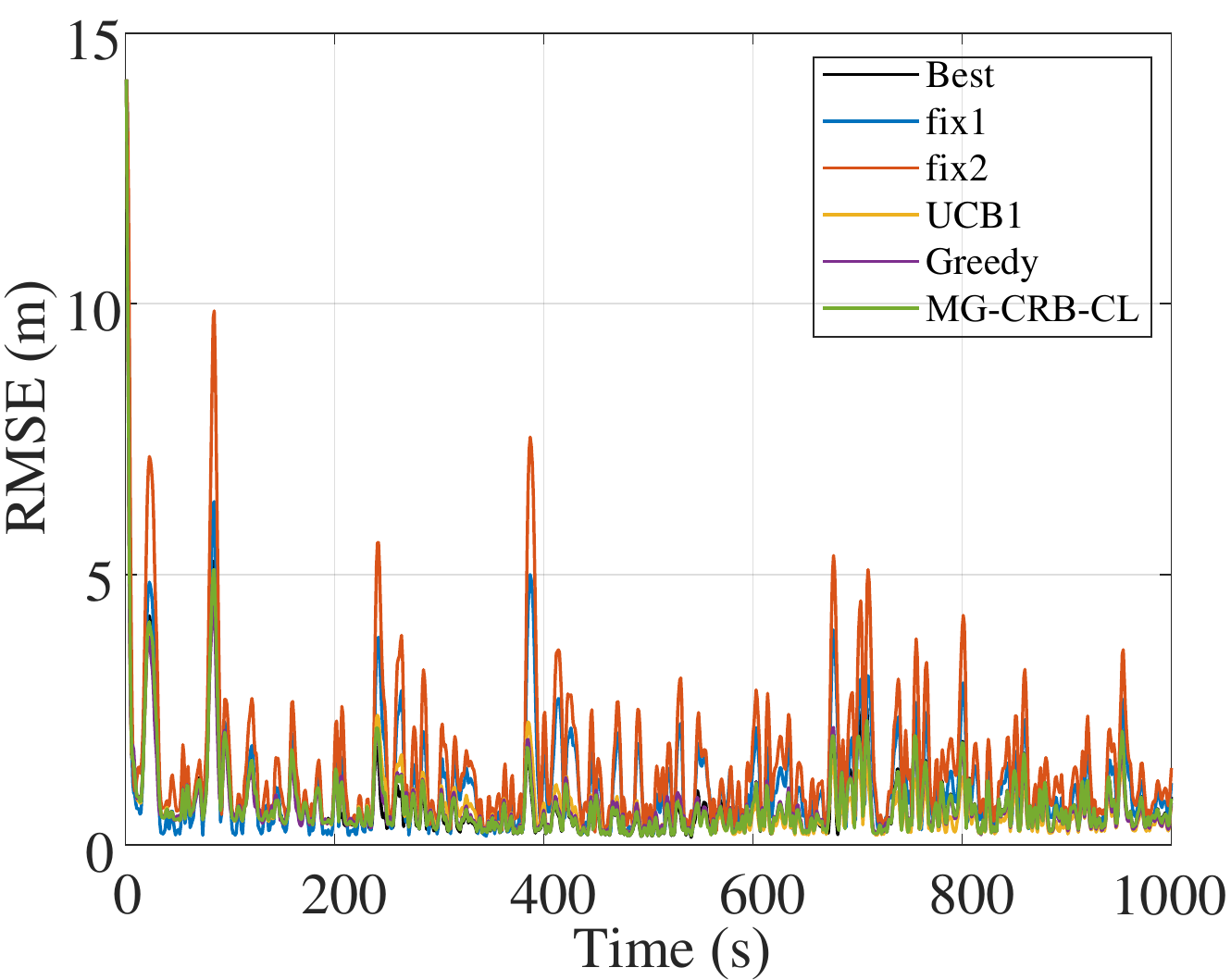}
        \centerline{(c)}
    \end{minipage}
    \caption{RMSE (m) for each algorithm in Scenario 3. (a) RMSE (m) for targets. (b) RMSE (m) of Target 1. (c) RMSE (m) of Target 2.}
    \label{fig_sim18}
\end{figure*}

The RMSE of the weighted fusion on the tracking performance is plotted in Fig.~\ref{fig_sim18}(a). Fig.~\ref{fig_sim18}(b) and~\ref{fig_sim18}(c) provide the RMSE results for Target 1 and Target 2, respectively. 

All in all, it can be concluded that MG-CRB-CL clearly outperforms the algorithms with fixed selections (that is, UCB1 and $\epsilon$-greedy) with respect to the multi-target RMSE.
It validates the advantages of taking into account the dynamic resource scheduling and is consistent with the SINR results plotted in Fig.~\ref{fig_sim16}. 
The results for ARMSE are presented in Table~\ref{tab4}, where MG-CRB-CL achieves the lowest RMSE on Target 1 and compatible RMSE on Target 2. 
Because of the predefined weight $\Omega_k$ of target $k$ in the optimization objective function in~\eqref{eq17}, UCB1 obtains better RMSE for Target 2 than that of MG-CRB-CL.
However, MG-CRB-CL outperforms UCB1 with respect to multi-target RMSE, validating the near-optimality of MG-CRB-CL.
\begin{table}[!t]
\begin{center}
\caption{Comparison of ARMSE (m) through the Time Horizon \textit{T}}
\label{tab4}
\begin{tabular}{c | c | c | c}
\hline
{Algorithms} & {ARMSE} & {ARMSE} & {ARMSE}\\ 
 & {of Target 1} & {of Target 2} & {of targets}\\
\hline
Best & 0.7539 & 0.7543 & 0.7541 \\
\hline
MG-CRB-CL & 0.7649 & 0.7527 & 0.7588\\
\hline
UCB1 & 0.9329 & 0.7145 & 0.8237\\
\hline
$\epsilon$-greedy & 0.7873 & 0.7478 & 0.7675 \\
\hline
fix1 & 0.8575 & 1.1237 & 0.9906\\
\hline
fix2 & 1.2724 & 1.5643 & 1.4184\\ 
\hline 
\end{tabular}
\end{center}
\end{table}

\begin{figure}[!t]
    \begin{minipage}[t]{0.5\linewidth}
        \centering
        \includegraphics[width=1.55in]{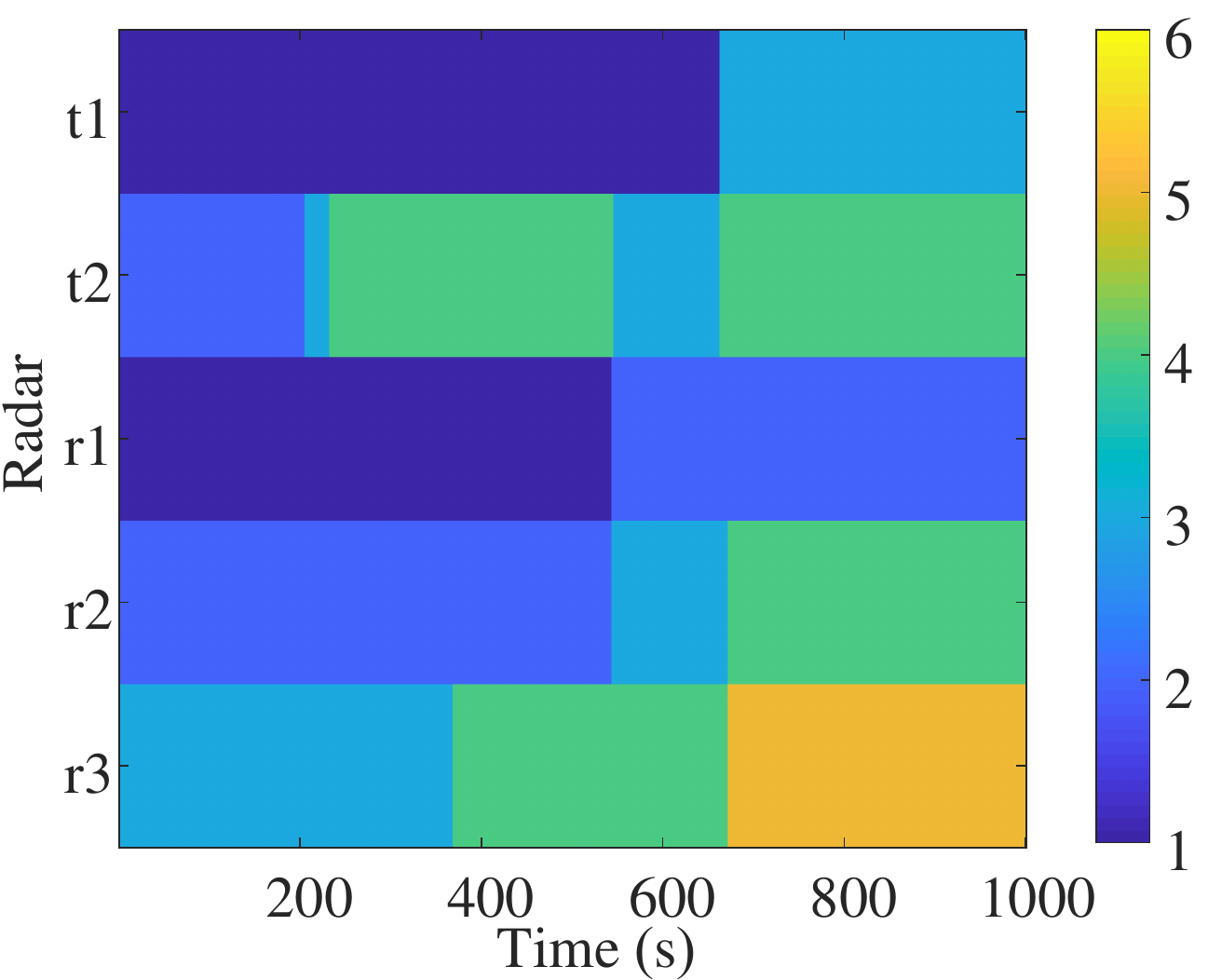}
        \centerline{(a)}
    \end{minipage}%
    \begin{minipage}[t]{0.5\linewidth}
        \centering
        \includegraphics[width=1.55in]{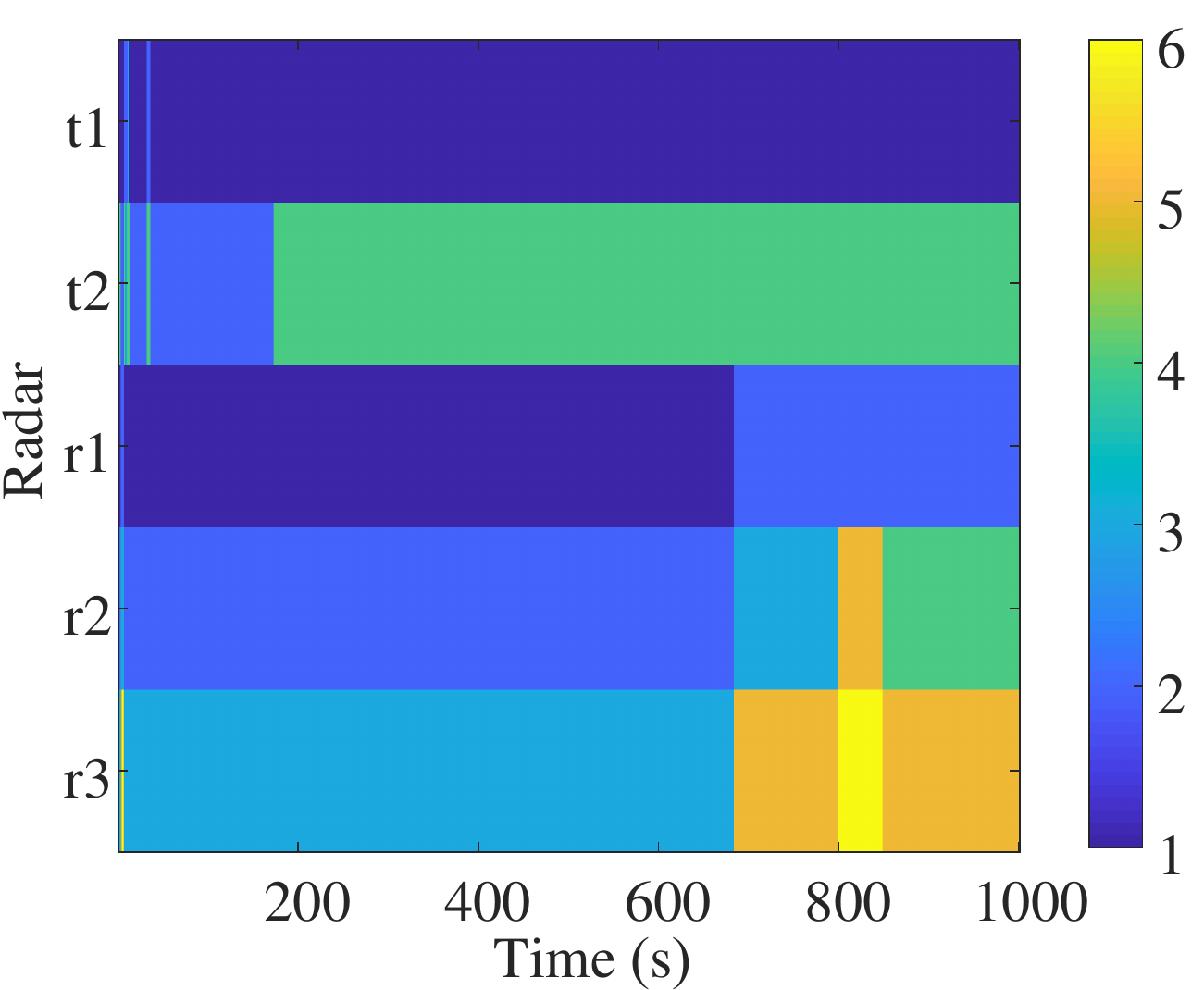}
        \centerline{(b)}
    \end{minipage}
    \begin{minipage}[t]{0.5\linewidth}
        \centering
        \includegraphics[width=1.55in]{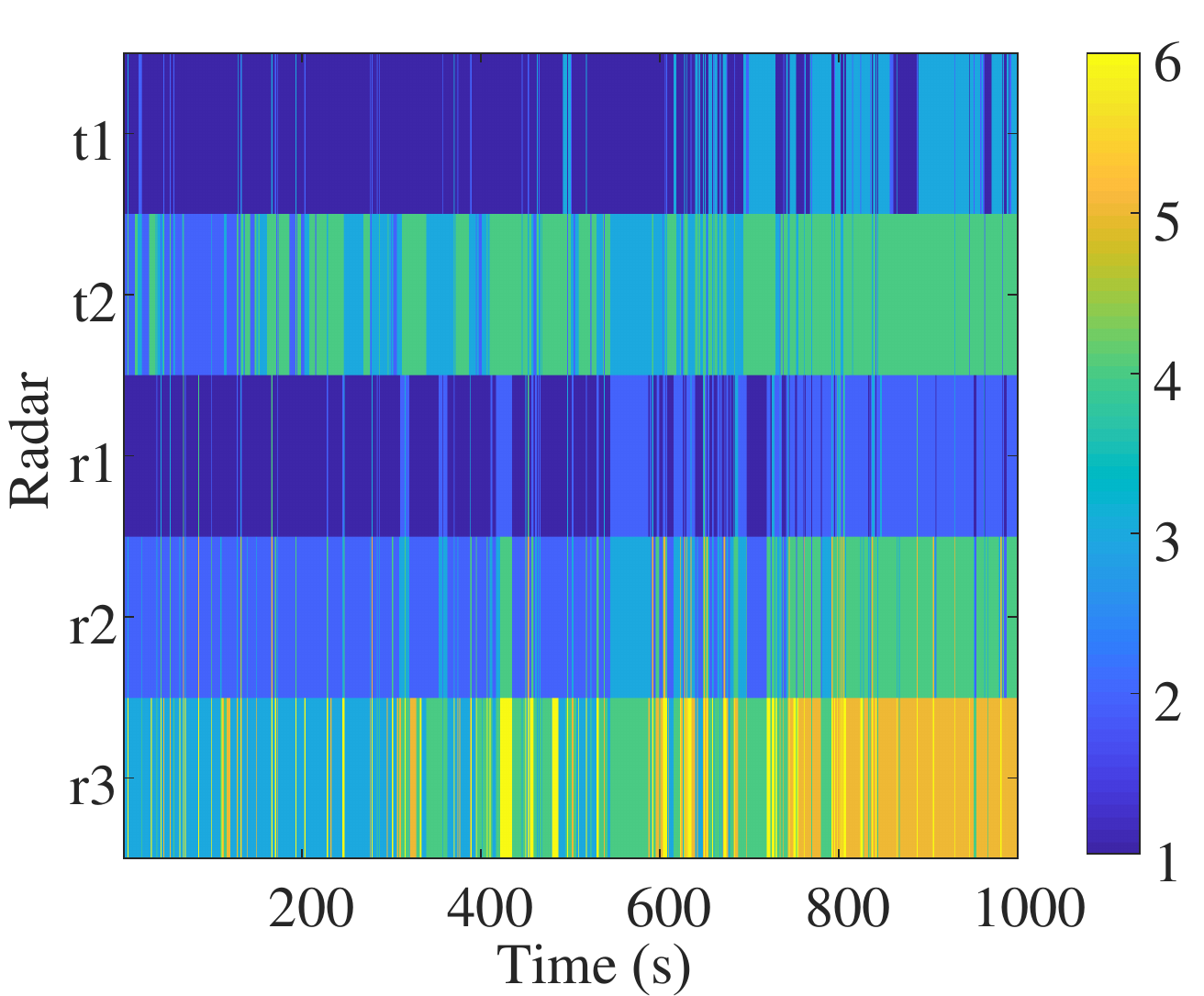}
        \centerline{(c)}
    \end{minipage}%
    \begin{minipage}[t]{0.5\linewidth}
        \centering
        \includegraphics[width=1.55in]{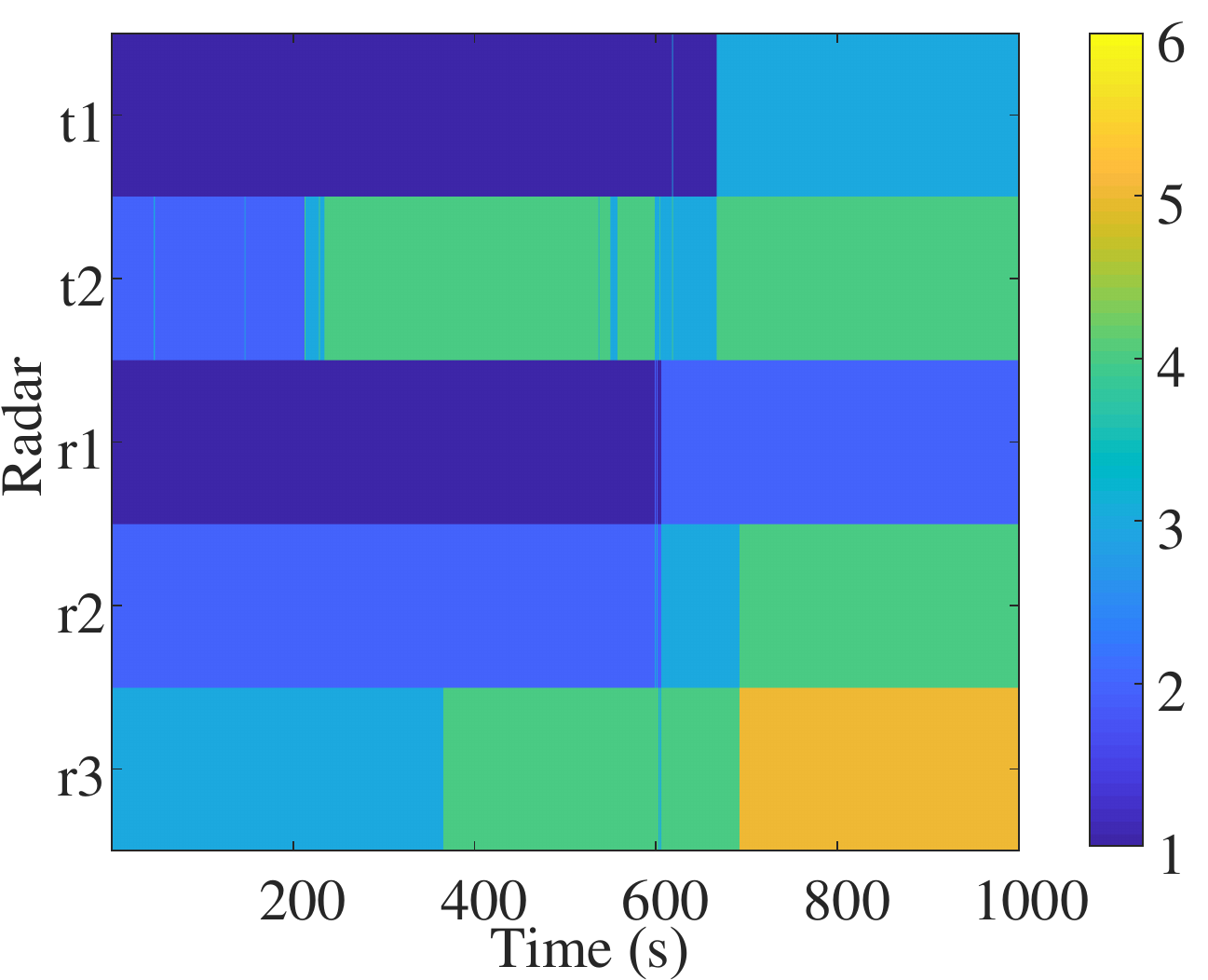}
        \centerline{(d)}
    \end{minipage}%
    \caption{Radar station selection index in time. (a) Best policy. (b) UCB1 algorithm. (c) $\epsilon$-greedy algorithm. (d) MG-CRB-CL algorithm.}
    \label{fig_sim21}
\end{figure}

The selection results are shown in Fig.~\ref{fig_sim21} with color maps, for which 
MG-CRB-CL has the highest ASR, $69.08\%$, and selects the same radar stations as the Best policy in most of the time.
While UCB1 and $\epsilon$-greedy obtain only $17.94\%$ and $31.56\%$ as the average rates, respectively.

The three simulations numerically demonstrate the effectiveness and near-optimality of MG-CRB-CL in both single-target and multi-target tracking cases. 

\section{Conclusion}\label{sec:conclusion}
We have formulated the TX-RX selection problem as an MG-CRB process that considers moving targets and non-stationary channels for distributed MIMO radars. 
It follows a trade-off between the exploration of unknown rewards of channels and the exploitation of the best TX-RX pairs based on current estimations.
We have proposed the MG-CRB-CL algorithm that aims to enhance the accuracy of target tracking, where IMM-UKF has been adapted to estimate target locations and a closed-loop framework has been established to approximate SINR in the next time slot.
We have also adapted the BPSO method to select the best super arm, which is an important step for the MG-CRB-CL algorithm.
Through extensive simulation results for both non-maneuvering and maneuvering scenarios, we have demonstrated the effectiveness of the proposed MG-CRB-CL algorithm by comparing it to the benchmark algorithms. 
For future research, we plan to incorporate joint power and radar selection optimization in the RMAB model and adapt the MG-CRB-CL algorithm to the new case.


%




\ifCLASSOPTIONcaptionsoff
  \newpage
\fi



%

\bibliographystyle{IEEEtran}
\bibliography{IEEEabrv,reference}

\end{document}